\documentclass[twocolumn,preprintnumbers,amsmath,amssymb]{revtex4-1}
\usepackage{graphicx}% Include figure files
\usepackage{ulem}
\usepackage[svgnames]{xcolor}
\usepackage{bm}% bold math
\usepackage{multirow}
\newif\ifshowcomments\showcommentstrue
\begin{document}

\title{Finite-temperature violation of the anomalous transverse Wiedemann-Franz law}

\author{ Liangcai Xu$^{1}$,  Xiaokang Li$^{1,2}$, Xiufang Lu$^{1}$, Cl\'ement Collignon$^{2,3}$,  Huixia Fu$^{4}$, Jahyun Koo$^{4}$ , Beno\^{\i}t Fauqu\'e$^{2,3}$}
\author{Binghai Yan$^{4,}$}\email{binghai.yan@weizmann.ac.il}
\author{ Zengwei Zhu$^{1,}$}\email{zengwei.zhu@hust.edu.cn}
\author{Kamran Behnia$^{1,2,5,}$}\email{kamran.behnia@espci.fr}

%Zengwei Zhu$^{1,*}$ and Kamran Behnia$^{1,2,5,*}$}

\affiliation{(1) Wuhan National High Magnetic Field Center and
 School of Physics, Huazhong University of Science and Technology,  Wuhan  430074, China\\
 2) Laboratoire de Physique et d'Etude des Mat\'{e}riaux (CNRS)\\ ESPCI Paris, PSL Research University, 75005 Paris, France\\
(3) JEIP, USR 3573 CNRS, Coll\`ege de France, 11 place Marcelin Berthelot, 75005 Paris, France\\
(4) Department of Condensed Matter Physics, Weizmann Institute of Science, 7610001 Rehovot, Israel\\
(5) II. Physikalisches Institut, Universit\"{a}t zu K\"{o}ln, 50937 K\"{o}ln, Germany}

\date{\today}
\begin{abstract}
The Wiedemann-Franz (WF) law links the ratio of electronic charge and heat conductivity to fundamental constants. It has been tested in numerous solids, but the extent of its relevance to the anomalous transverse transport, which represents the topological nature of the wave function, remains an open question. Here we present a study of anomalous transverse response in the noncollinear antiferromagnet Mn$_{3}$Ge extended from room temperature down to sub-Kelvin temperature and find that the anomalous Lorenz ratio remains close to the Sommerfeld value up to 100 K, but not above. The finite-temperature violation of the WF correlation is caused by a mismatch between the thermal and electrical summations of the Berry curvature, rather than the inelastic scattering as observed in ordinary metals. This interpretation is backed by our theoretical calculations, which reveals a competition between the temperature and the Berry curvature distribution.
The accuracy of the experiment is supported by the verification of the Bridgman relation between the anomalous Ettingshausen and Nernst effects. Our results identify the anomalous Lorenz ratio as an extremely sensitive probe of Berry spectrum near the chemical potential.
\end{abstract}
\maketitle

\section*{Introduction}
The Berry curvature of electrons can give rise to the anomalous Hall effect (AHE)\cite{Nagaosa2010,Xiao2010}. This happens if the host solid lacks time-reversal symmetry, which impedes cancellation after integration over the whole Fermi surface. Explored less frequently\cite{Lee2004,Miyasato2007,Onose2008}, the thermoelectric and thermal counterparts of the AHE (the anomalous Nernst and anomalous Righi-Leduc effects) arise also by the same fictitious magnetic field\cite{Xiao2006,Onoda2008, Shiomi}. How do the magnitudes of these anomalous off-diagonal coefficients correlate with each other? Do the established correlations between the ordinary transport coefficients hold? Satisfactory answers to these questions are still missing. A  semiclassical formulation of anomalous Hall effect\cite{Sinitsyn2008} is laborious, because the concept of Berry connection (or the `anomalous velocity' \cite{Karplus1954}) is based on off-diagonal matrix elements linking adjacent Bloch functions and not wave packets of semi-classical transport theory\cite{Sinitsyn2008}. This makes any intuitive picture of how Berry curvature combined to a longitudinal thermal gradient can produce a transverse electric field (an anomalous Nernst response)\cite{Xiao2006,Mccormick} or a  transverse thermal gradient (an anomalous thermal Hall response)\cite{Murakami,Qin} even more challenging.

Here, we present a study of correlations between the anomalous off-diagonal transport coefficients of a magnetic solid, with a focus on the relation between  anomalous electrical, $\sigma^{A}_{ij}$, and thermal, $\kappa^{A}_{ij}$, Hall conductivities. The anomalous Lorenz ratio is defined as:

\begin{equation}\label{1}
L^{A}_{ij}=\frac{\kappa^{A}_{ij}}{T\sigma^{A}_{ij}}.
\end{equation}
We track $L^{A}_{ij}$ in order to compare it with the Sommerfeld value:
\begin{equation}\label{2}
 L_0=\frac{\pi^2}{3}\left(\frac{k_B}{e}\right)^2.
\end{equation}
We find that over a wide temperature range ($0.5$ K $ < T < 100$ K), $L^{A}_{ij}$ and $L_0$ remain close to each other, but a deviation starts above 100 K. We argue that this observation implies a hitherto unnoticed mechanism for finite-temperature violation of the Wiedemann-Franz (WF) law and points to a small (10 meV) energy scale in the Berry spectrum of Mn$_{3}$Ge that is absent in Mn$_{3}$Sn. This experimental observation is backed by theoretical calculations identifying  the source of the Berry curvature in this family. By directly measuring the anomalous Ettingshausen and anomalous Nernst effects, we verify the validity of the Bridgman relation connecting the two transverse thermoelectric coefficients to each other. This confirms that the Bridgman relation, a consequence of Onsager reciprocity and based on thermodynamics of irreversible processes\cite{Callen}, holds regardless of microscopic details. Finally, we quantify the anomalous transverse thermoelectric response $\alpha^{A}_{ij}$ and find that the ratio of $\alpha^{A}_{ij}/\sigma^{A}_{ij}$ tends to saturate towards a value close to $k_B/e$ in the high-temperature limit.

Following theoretical propositions\cite{Chen2014,Kubler2014}, a large AHE was found in Mn$_{3}X$ ($X=$ Sn, Ge) family of non-collinear antiferromagnets \cite{Nakatsuji2015,Nayak2016,Kiyohara2016}  below a remarkably high N\'{e}el temperature\cite{Zimmer1973,Tomiyoshi1982,Tomiyoshi1982b}. These newcomers to the emerging field of antiferromagnetic spintronics\cite{Smejkal2018}, present a distinct profile of the Hall resistivity in which the extraction of the anomalous Hall conductivity becomes straightforward. An anomalous thermoelectric (Nernst)\cite{Ikhlas2017,Li2017} and Righi-Leduc\cite{Li2017}, counterparts of AHE, were also observed in Mn$_3$Sn. In the case of Mn$_3$Sn the triangular order is destroyed at finite temperature\cite{Nakatsuji2015,Ikhlas2017,Li2017}. This is not the case of Mn$_3$Ge where the fate of these signals can be followed down to sub-Kelvin temperatures.
\begin{figure}
\includegraphics[width=9cm]{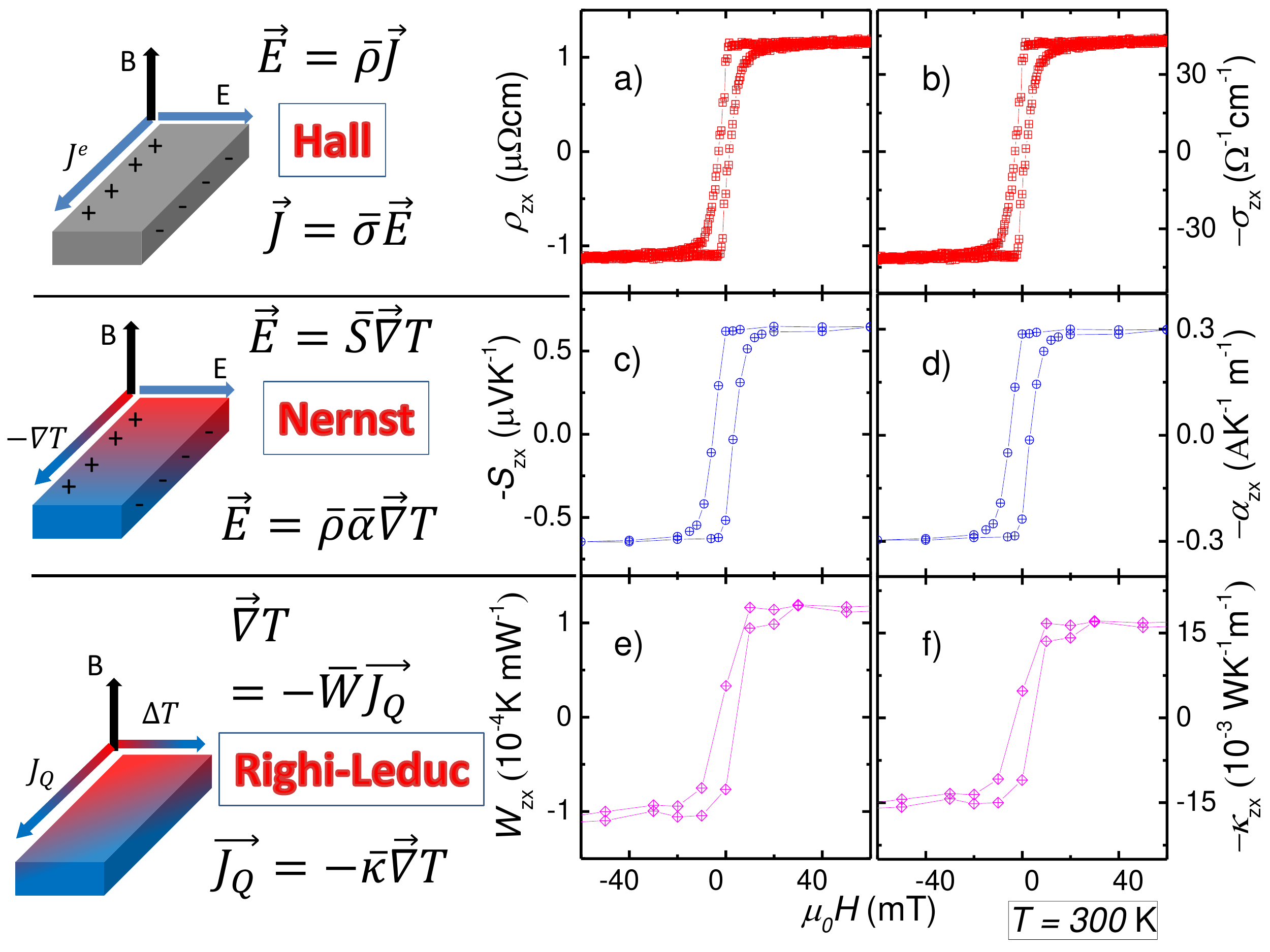}
\caption{\textbf{Anomalous transverse coefficients, definitions and profiles:} Off-diagonal components of the three conductivity tensors (electric, $\bar{\sigma}$, thermoelectric, $\bar{\alpha}$, and thermal, $\bar{\kappa}$) can be finite in the absence of magnetic field.  As shown in the three left panels, they  link to four vectors, which are charge density current, $\vec{J}$, electric field, $\vec{E}$, thermal gradient, $\vec{\nabla} T$ and heat density current, $\vec{J_Q}$. a) Hall resistivity, $\rho_{zx}$; b) Hall conductivity, $\sigma_{zx}$,  extracted from $\rho_{zx}$, $\rho_{xx}$ and $\rho_{zz}$; c) Nernst signal, $S_{zx}$; d) Transverse thermoelectric conductivity, $\alpha_{zx}$, extracted from $S_{zx}$, $S_{xx}$, $\rho_{xx}$, $\rho_{zz}$ and $\rho_{zx}$; e) thermal Hall resistivity, $W_{zx}$; f) thermal Hall conductivity, or the Righi-Leduc coefficient, $\kappa_{zx}$ extracted from off-diagonal and diagonal thermal resistivities.
\label{fig:anom-trans-resp}}
\end{figure}
\begin{figure}
\includegraphics[width=8cm]{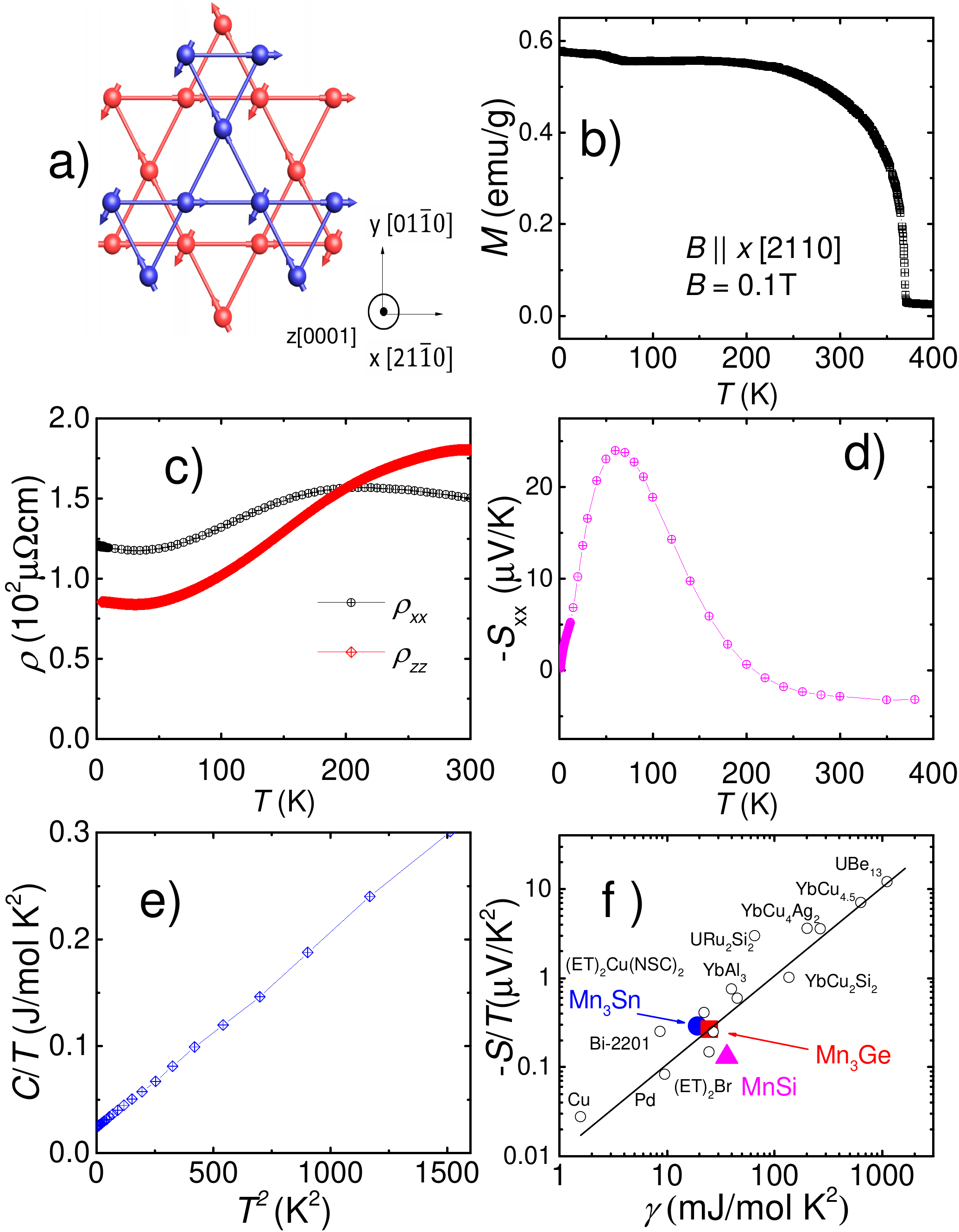}
\caption{\textbf{Antiferromagnetic, dirty and correlated:} a) A sketch of the magnetic texture of Mn$_3$Ge showing the orientation of spins of Mn atoms. Red and blue represent two adjacent planes. b) Temperature dependence of the magnetization with N\'eel temperature visible at 370 K. c) Temperature dependence of resistivity along two orientations. d) The Seebeck coefficient, $S$, as a function of temperature. e) Low temperature specific heat\, $C/T$, as a function of $T^2$. Extrapolation to $T=0$ yields $\gamma = 24.3$ mJ mol$^{-1}$K$^{-2}$. f) Plot of the absolute value of $S/T$ vs. $\gamma$ for a number of correlated metals including Mn$_3$X and MnSi\cite{Stishov2008,Stevan2013}.}
\label{fig:basic-prop}
\end{figure}

\section*{Basic properties}
The room-temperature field-dependence of the three transport properties in  Mn$_3$Ge is shown in Fig.~\ref{fig:anom-trans-resp}. Like in Mn$_3$Sn, a hysteretic jump is triggered at a well-defined magnetic field, marking the nucleation of domains of opposite polarity induced by magnetic field \cite{Li2018}. The large jump, the small magnetic field required for inverting polarity and the weakness of the ordinary Hall response lead to step-like profiles contrasting with other topological solids exhibiting AHE \cite{Liang2017,Liang2018,Shekhar2018}. A step-like profile of anomalous transverse response (for other varieties\cite{Sakai2018,Liu2018}) makes the extraction of the anomalous component straightforward. Panels of Fig.~\ref{fig:anom-trans-resp} show the measured Hall resistivity, Nernst signal and thermal Hall resistivity, which were used to extract electric, thermoelectric and thermal Hall conductivities.

Figure~\ref{fig:basic-prop} presents a number of basic properties of the system under study. The spin texture\cite{Zimmer1973,Tomiyoshi1982,Tomiyoshi1982,Nakatsuji2015} is shown in Fig.~\ref{fig:basic-prop}a. This magnetic order is stabilized thanks to the combination of Heisenberg and Dzyaloshinskii-Moriya interactions\cite{Liu2017}. As seen in Fig.~\ref{fig:basic-prop}b, which shows the magnetization, it emerges below $T_N = 370$ K. The small residual ferromagnetism  has been attributed to the residual magnetic moment of octupole clusters of Mn atoms\cite{Suzuki2017} in this pseudo-Kagome lattice.

A carrier density of $n=3.1\times10^{22}$ cm$^{-3}$ is extracted from the magnitude of the ordinary Hall number\cite{supplement} in agreement with a previous report\cite{Kiyohara2016}. The electrical resistivity shows little variation with temperature (Fig.~\ref{fig:basic-prop}c).  and its magnitude of 150 $\mu\Omega$cm implies a  mean free path as short as 0.9 nm, compatible with the fact that Mn$_{3}$X crystals are not stoichiometric\cite{Tomiyoshi1982,Kiyohara2016}. In our crystals, we found the Mn:Ge ratio ranges from 3.32:1 to 3.35:1\cite{supplement}. Since one tenth of Ge sites are occupied by Mn atoms, the average distance between these defects is $\sqrt[3]{10}$ times the average lattice parameter ($a=0.53$ and $c=0.43$ nm) and comparable to the short electronic mean free path.
\begin{figure*}
\centering
\includegraphics[width=16cm]{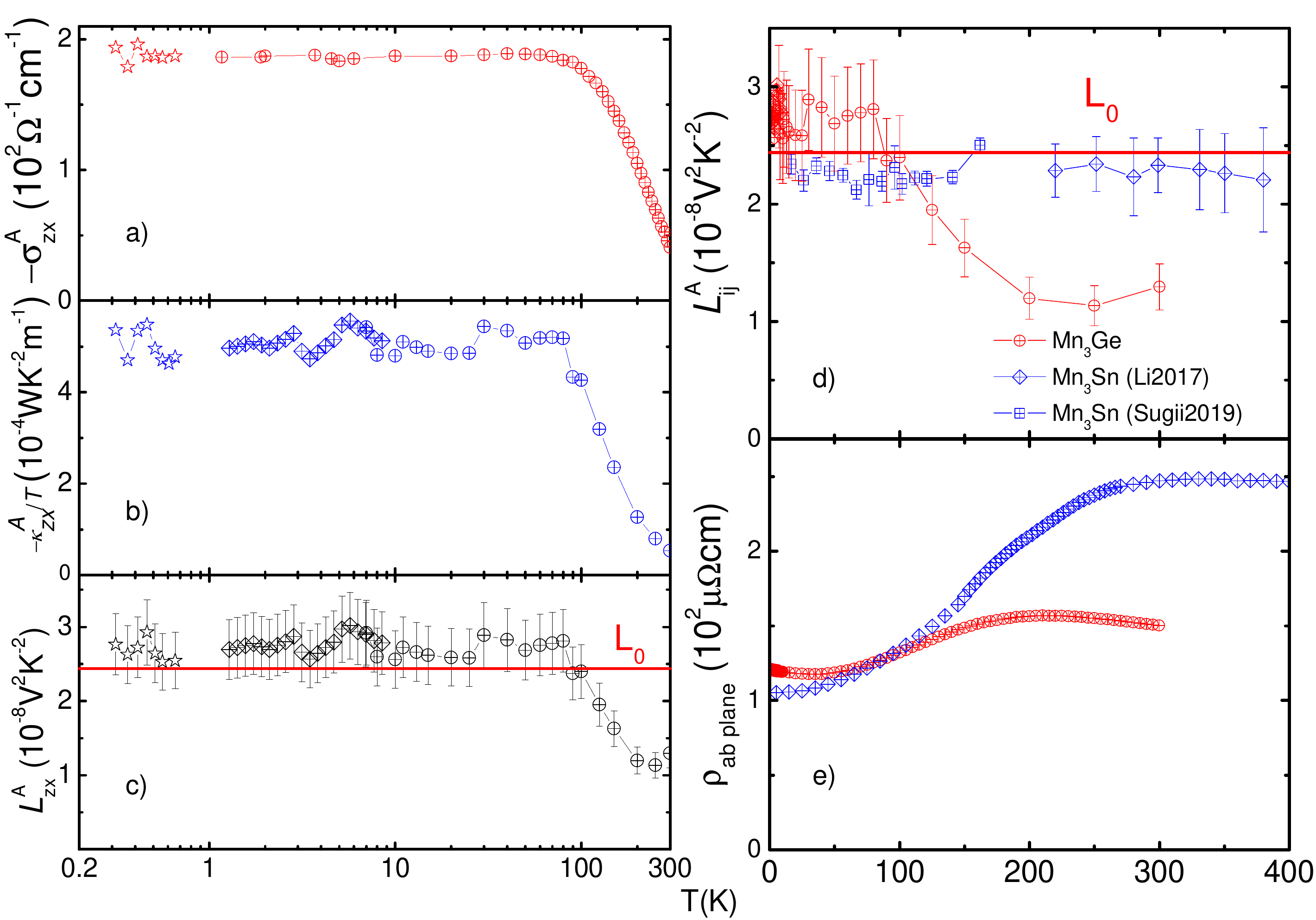}
\caption{\textbf{Anomalous transverse Wiedemann-Franz law:} Temperature dependence of the anomalous Hall conductivity $\sigma^{A}_{zx}$ (a) ; the anomalous thermal Hall conductivity divided by temperature $\kappa^A_{zx}/T$  (b); c) the anomalous Lorenz ratio $\kappa^A_{zx}/\sigma^{A}_{zx}T$. Different symbols are used for data obtained with two different set-ups: resistive thermometers (diamonds) and thermocouples (circles). Star symbols refer to a third set of data obtained on another sample measured  down to sub-Kevin temperatures. The horizontal solid line marks $L_{0}=2.44 \times 10^{-8}$ V$^{2}$K$^{-2}$. The deviation between $L$ and $L_0$ starts at $T>100$ K and is concomitant with the decrease in $\sigma^{A}_{zx}$.
 d) Temperature dependence of the anomalous Lorenz ratio in Mn$_3$Ge and in Mn$_3$Sn\cite{Li2017,Sugii}. e) Comparison of their in-plane resistivity. The large deviation from the WF law in Mn$_3$Ge occurs in spite of the fact that the temperature dependence of its resistivity is even more modest than that in Mn$_3$Sn.
\label{fig:anom-wf}}
\end{figure*}

The Seebeck coefficient (Fig.~\ref{fig:basic-prop}d) has a nonmonotonic temperature dependence with a peak around 60 K, and a sign change above 200 K, and a large low-temperature slope indicative of electronic correlations. The $T$-linear electronic specific heat (Fig.~\ref{fig:basic-prop}e), is as large as $\gamma=24.3$ mJ mol$^{-1}$K$^{-2}$, thirty times larger than copper and five times larger than iron\cite{Kittel}. Assuming a single spherical Fermi surface corresponding to the known carrier density, such a $\gamma$, implies an effective mass as large as $m^* = 14.5\,m_e$, which should not be taken literally given that the system is multiband. The slope of the Seebeck coefficient at low-temperature ($S/T \simeq -0.2~\mu$VK$^{-2}$) correlates with $\gamma$ yielding $q=\frac{SN_{Av}e}{T\gamma}\simeq1 $, as observed in other correlated systems\cite{Behnia2004} (Fig.~\ref{fig:basic-prop}f). On the other hand, the low temperature resistivity presents a weak upturn impeding the detection of any $T$-square term and the Wilson ratio (of the specific heat and magnetic susceptibility \cite{Coleman2007}) implies that mobile electrons do not play any significant role in the magnetic response\cite{supplement}.

\section*{The anomalous transverse WF law}
For each temperature, we measured $\sigma^{A}_{zx}$ and $\kappa^A_{zx}$, identified as jumps in $\sigma_{zx}(B)$ and  $\kappa_{zx}(B)$. This led to the determination of $L^{A}_{zx}$ at each temperature and a comparison with $L_0$ to check the WF law.

Our main finding is presented in Fig.~\ref{fig:anom-wf}. Below 100 K, the anomalous Lorenz ratio, $L^A_{zx}$ was found to be flat with a magnitude slightly larger than the Sommerfeld value, equal to it within the experimental margin. The results were reproduced by two different measuring methods (with resistive sensors and thermocouples\cite{supplement}) and in two different samples. In one of them, we checked the persistent validity of this equality below 1 K down to 0.3 K. As seen also in the figure, $L^A_{zx}$ begins a steady downward deviation from ${L_0}$ above 100 K. Interestingly, $\sigma^{A}_{zx}$ begins a steady decrease itself above 100 K. The temperature dependence of $\sigma^{A}_{zx}$ is similar to what was reported in previous reports\cite{Nayak2016,Kiyohara2016} and its zero-temperature magnitude (which depends on stoichiometry\cite{Kiyohara2016}) is in agreement with was reported for a similar Mn content \cite{Kiyohara2016, supplement}.
As seen in Fig.~\ref{fig:anom-wf}d, the  anomalous Lorenz ratio in Mn$_3$Ge  and in Mn$_3$Sn behave very  differently, in spite of the fact that resistivity in both shows only a slight change with temperature (Fig.~\ref{fig:anom-wf}e), in contrast to elemental ferromagnets.

\begin{figure}
\includegraphics[width=8cm]{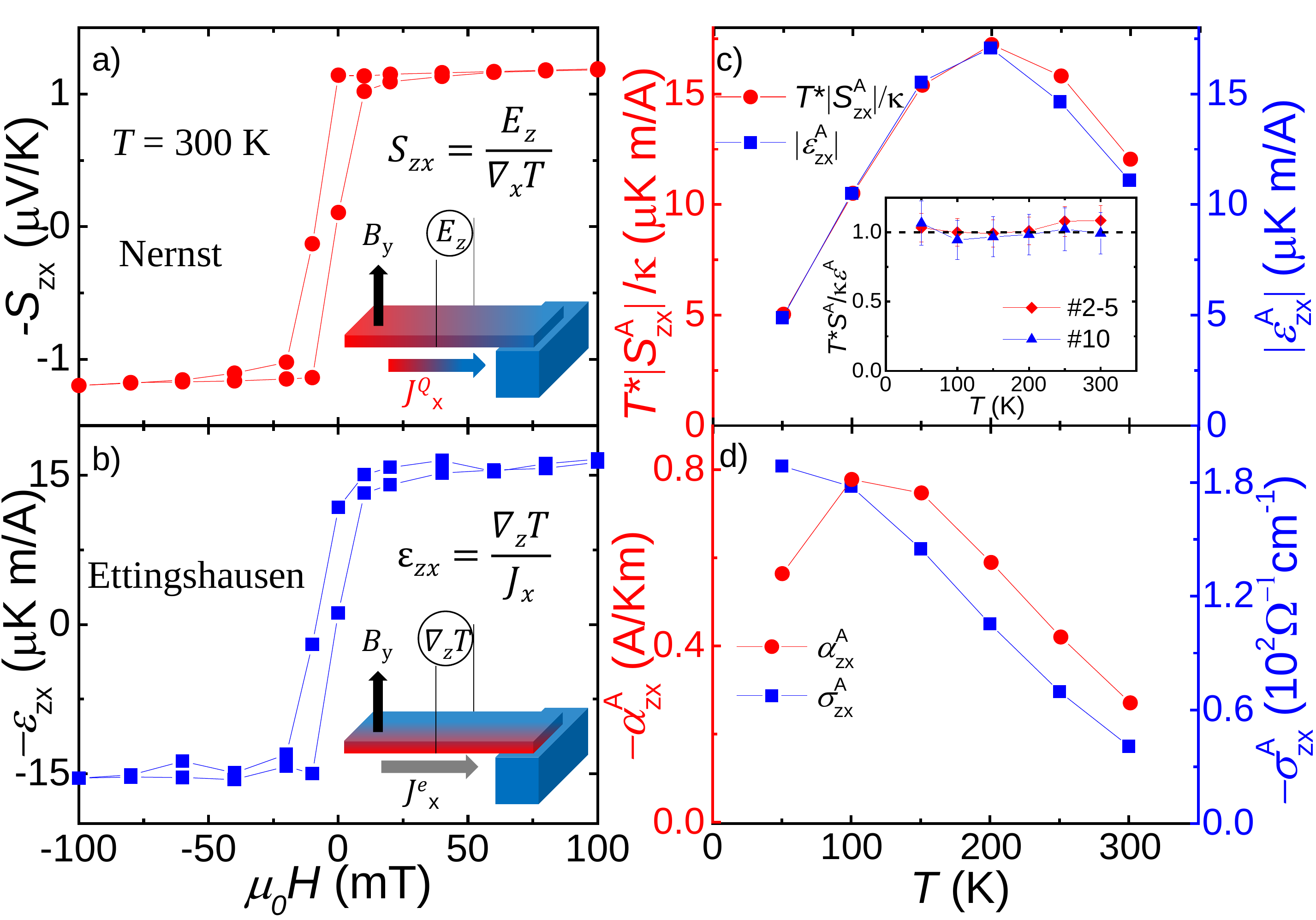}
\caption{\textbf{Anomalous Nernst and Ettingshausen effects and the Bridgman relation:} a) The transverse electric field created by a finite longitudinal temperature gradient as a function of magnetic field (the Nernst effect); b) the transverse thermal gradient produced by a finite longitudinal charge current (the Ettingshausen effect) at the same temperature. Insets show experimental configurations. c) The temperature dependence of the anomalous Nernst ($S^A_{zx}$) and anomalous Ettingshausen ($\epsilon^A_{zx}$) coefficients. $\epsilon^A_{zx}$ and $S^A_{zx}T/\kappa_{xx}$ remain equal as expected by the Bridgman relation. d) Temperature dependence of $\alpha^A_{zx}$ extracted from the Nernst signal $S^A_{zx}$. Also shown is the temperature dependence of Anomalous Hall conductivity $\sigma^A_{zx}$.
\label{fig:anom-nern-ett-brid}}
\end{figure}

\section*{The Bridgman relation}
Several previous reports of the violation of WF law have been refuted afterwards. One may therefore wonder if our data  can be validated by an independent criteria. The answer is affirmative. Their validity is supported by the verification of the Kelvin relation (for normal longitudinal transport coefficients) and the Bridgman relation (for anomalous transverse coefficients). Their violation  counter the thermodynamics of irreversible processes.

To check the Bridgman relation, we directly measured both the Nernst (Fig.~\ref{fig:anom-nern-ett-brid}a) and the Ettingshausen (Fig.~\ref{fig:anom-nern-ett-brid}b) effects. The former is the transverse electric field generated by a longitudinal thermal gradient, $S_{zx}= \frac{E_{z}}{\nabla_{x}T}$ and the latter is the transverse thermal gradient produced by a longitudinal charge current, $\epsilon^{A}_{zx}= \frac{\nabla_{z}T}{J_x}$.  The anomalous Ettingshausen effect (Fig.~\ref{fig:anom-nern-ett-brid}b) is easily invertible by a small magnetic field, like other anomalous transverse responses. We measured  $S^{A}_{zx}$ and $\epsilon^{A}_{zx}$ at several different temperatures. Combined with longitudinal thermal conductivity data $\kappa_{xx}(T)$, this allowed us to check the Bridgman relation\cite{Bridgman}, which links these three quantities:

\begin{equation}\label{3}
\epsilon^{A}_{zx}=\frac{T S^{A}_{zx}}{\kappa_{xx}}.
\end{equation}

As seen in Fig.~\ref{fig:anom-nern-ett-brid}c, the two sides of the equation remain close to each other in the whole temperature range. The Bridgman relation, derivable by a thermodynamic argument\cite{Sommerfeld} is based on Onsager reciprocity\cite{Callen}. Its experimental validity has been confirmed in  semiconductors\cite{Delves} and in superconductors hosting mobile vortices\cite{Huebener}. While there is a previous report on simultaneous measurements of anomalous Nernst and Ettingshausen coefficients\cite{Seki},  the present study is the first experimental confirmation of the validity of Bridgman relation in the context of topological transverse response. We also verified the Kelvin relation linking the Seebeck and Peltier coefficients\cite{supplement}.

The temperature dependence of the anomalous transverse thermoelectric conductivity, $\alpha^A_{zx}$, extracted from anomalous Nernst coefficient, is presented in Fig.4d. As expected, it vanishes in the zero-temperature limit, but is remarkably large at room temperature. This can be seen by noting that the ratio of $\alpha^A_{zx}$ to $\sigma^A_{zx}$ is close to $k_B/e$ at room temperature. This contrasts  with the ordinary longitudinal counterpart of this ratio which includes a $T/T_F$ damping factor, inversely scaling with the Fermi temperature, $T_F$\cite{Behnia2016}.

\begin{figure}
\includegraphics[width=8.5cm]{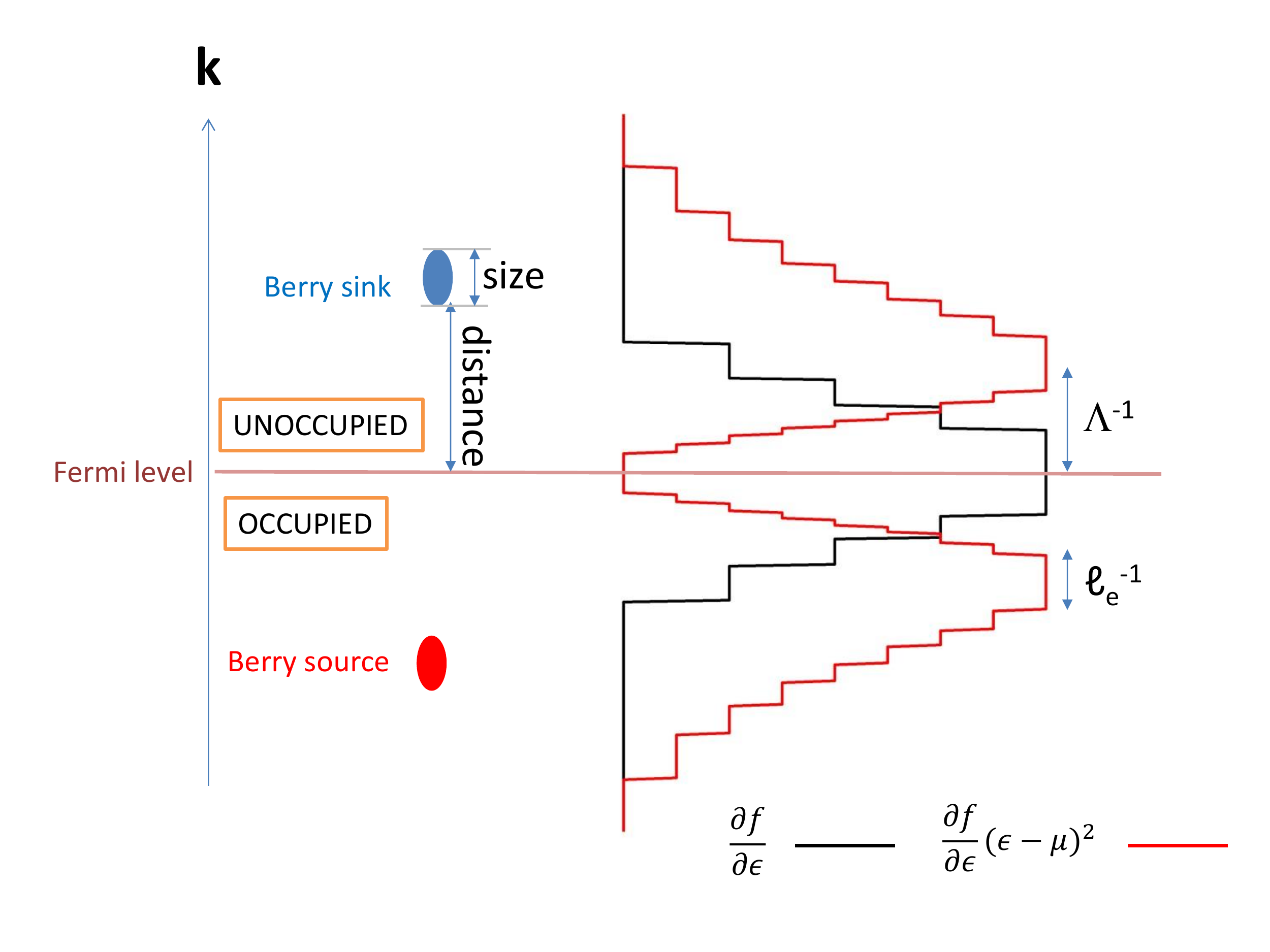}
\caption{\textbf{Thermal and electrical transport  and length scales in $k$-space:} The pondering factor for charge transport  ($\partial f(\epsilon_k)/\partial \epsilon_k$) is centered at the Fermi level. The same function for thermal transport ($(\epsilon_k - \mu)^{2} \partial f(\epsilon_k)/ \partial \epsilon_k$) has two peaks shifted off the Fermi level.  A mismatch between thermal and electrical summations can occur if the two pondering functions differently average the overall contribution of the Berry curvature sources and sinks. Disorder makes these functions step-like in $k$-space. Note four distinct relevant length scales in the $k$-space.
\label{fig:contrast}}
\end{figure}

\section*{Origin of the finite-temperature violation}
Having discussed the thermoelectric response, let us now turn back to the thermal transport. The zero-temperature validity of the WF law implies that the transverse flow of charge and entropy caused by Berry curvature conforms to a ratio of $\frac{\pi^{2}}{3}\left(\frac{k_{B}}{e}\right)^2$. This confirms Haldane's argument \cite{Haldane2004} that AHE is a property of the topological quasi-particles of the Fermi surface. As argued previously\cite{Li2017}, only the states within the thermal window of the Fermi surface can be affected by a temperature gradient and give rise to a finite $\kappa^{A}_{zx}$.  However, even in the case of ordinary longitudinal transport \cite{Ziman}, the WF law ceases to be valid in presence of inelastic scattering. This is because small-angle inelastic collisions decay the momentum flow less efficiently than the energy flow both for electron-phonon \cite{Ziman} and electron-electron\cite{Jaoui2018} scattering. However, if inelastic scattering played a significant role, one would have not observed such a drastic difference between Mn$_3$Sn and Mn$_3$Ge. Moreover, since the violation of the WF law is not correlated with any significant change in the magnitude of resistivity, it cannot be attributed to the gradual emergence of an extrinsic source of AHE \cite{Shiomi}.  Therefore, one should look for a  hitherto unidentified route to the finite-temperature violation of the anomalous transverse WF law.

In the semi-classic picture of electronic transport, charge and heat conductivity are set by the mean-free-path and the Fermi radius averaged over the whole Fermi surface with a pondering factor, which is\cite{vanhouten1992,Ziman,Behnia2015}:

\begin{equation}\label{4}
F_n(\epsilon_k)= (\epsilon_k-\mu)^{n}\frac{\partial f(\epsilon_k)}{\partial\epsilon_k}.
\end{equation}

This expression for the pondering factor has been obtained for both Boltzmann \cite{Ziman} and Landauer\cite{vanhouten1992} formalisms with $n=0$ for charge transport, $n= 1$  for thermoelectric transport and $n=2$ for thermal transport. Therefore, the main source  of each transport coefficient is not located at the same place in the $k$-space. The Berry curvature summed over the Fermi sheets with these pondering factors can potentially generate a mismatch between $\sigma^{A}_{ij}$ and $\kappa^{A}_{ij}$. As the temperature raises, the summations extend over a broader interval in the $k$-space inversely proportional to the thermal de Broglie length of electrons ($\Lambda=\frac{h}{\sqrt{2\pi m^*k_{B}T}}$)\cite{Behnia2015}. The electrical and thermal summations of the Berry curvature depend on these two length scales as well as the fine details of the spectrum, i.e. the size of the Berry curvature sources (and sinks) as well as their relative distance to the Fermi level (See Fig.~\ref{fig:contrast}). One can see that the validity of the WF correlation becomes harder to satisfy with warming. The higher the temperature, the larger the $k$-space area swept by the two different pondering functions and the easier to find a mismatch between  $\sigma^{A}_{ij}$ and $\kappa^{A}_{ij}$ if the Berry curvature changes significantly across $\Lambda^{-1}$. %On the other hand, because of disorder, the inverse of the mean-free-path, sets a minimum distance in $k$-space over which a Bloch wave is well-defined. %The thermal de Broglie length ($\Lambda$)  and the mean free path ($\ell_{e}$) can be both estimated. Because of the relative heaviness of electrons,  at 100 K, Interestingly, $\Lambda$ becomes as short as 1 nm and comparable to the mean free path ($\ell_{e}\sim$ 0.9 nm).
In this context, one can conceive a mismatch between thermal and electrical summations of the Berry curvature (whose typical scale in the momentum space is an order of magnitude smaller).  \color{black}

\section*{Theoretical calculations of the anomalous Lorenz ratio}
\vspace{1\baselineskip}

\begin{figure*}
 \centering
\includegraphics[width=18cm]{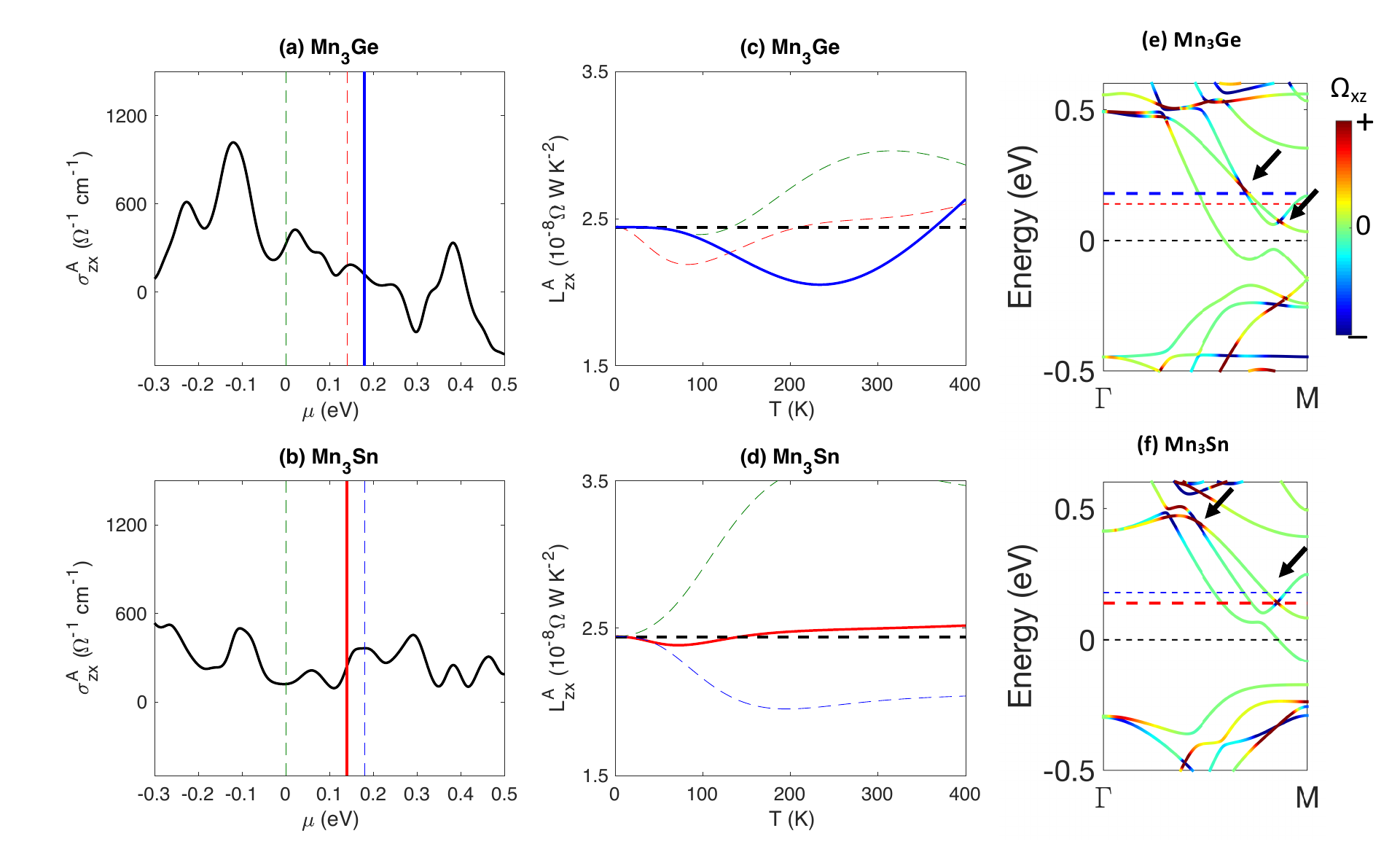}
\caption{The zero-temperature Berry curvature $\sigma_{zx}^A(\mu)$ (a) (c) and the anomalous Lorenz ratio $L^A_{zx}$ (c) (d). The charge neutral point is set to zero. The green, red and blue lines represent $\mu=$ 0, 140 and 180 meV, respectively. The dashed horizontal black lines represents $L_0$ in (c) and (d). In the band structure (e) and (f), the color indicates the Berry curvature value. The blue arrows point out two Weyl points between the lowest and second lowest conduction bands.
}
\label{calculation-Lxz}
\end{figure*}

The anomalous Hall conductivity $\sigma^A_{zx}$ and anomalous thermal Hall conductivity $\kappa^A_{zx}$ are expressed in the form of the Berry curvature $\Omega_{zx}^n(k)$ \cite{Vafek2001},

\begin{align}
    \sigma^A_{zx}(\mu)&= \frac{e^2}{\hbar}\int_{-\infty}^{\infty}d\xi \Bigg(-\frac{\partial f (\xi-\mu)}{\partial \xi}\Bigg)\tilde{\sigma}_{zx}(\xi),\label{simga}\\
    \kappa^A_{zx}(\mu)&= \frac{1}{\hbar T}\int_{-\infty}^{\infty}d\xi  \Bigg(-(\xi-\mu) ^2\frac{\partial f (\xi-\mu)}{\partial \xi}\Bigg)\tilde{\sigma}_{zx}(\xi),\label{kappa}\\
    \tilde{\sigma}_{zx}(\xi) &= \int_{BZ} \frac{d\textbf k}{(2\pi)^3}\sum_{\epsilon_n<\xi}\Omega^n_{zx}(\textbf k) \\
    \Omega^n_{zx}(\textbf k)&= \frac{1}{i} \sum_{m\ne n } \frac{\langle n|\hat{v}_z|m \rangle \langle n|\hat{v}_x|m \rangle - (x\leftrightarrow z) }{(\epsilon_n - \epsilon_m)^2}.
    \label{sigma2}
\end{align}

Here $f(\xi-\mu)=1/(e^\frac{\xi-\mu}{k_B T}+1)$ is the Fermi-Dirac function and $\hat{v}_{x/z}$ is the velocity operator,and $\tilde{\sigma}_{zx} (\xi)$ is actually the zero-temperature anomalous Hall conductivity (divided by $\frac{e^2}{\hbar}$). We point out that the electric and thermal coefficients are mainly different in the pondering factors for $\tilde{\sigma}_{zx} (\xi)$. As shown in Fig.~\ref{fig:contrast}, the pondering factor $ -\frac{\partial f (\xi-\mu)}{\partial \xi}$ in Eq.~\ref{simga} is a $\delta$-function while $ -(\xi-\mu)^2 \frac{\partial f (\xi-\mu)}{\partial \xi}$ in Eq.~\ref{kappa} displays a double-peak profile. The double-peak distance is about $5 k_B T$, where $T$ is the temperature. This is the essential cause to violate the anomalous WF law at finite $T$. It is clear that the anomalous Lorenz ratio $L^A_{zx} = \frac{\kappa^A_{zx}}{T\sigma^A_{zx}}$ remains as a constant $L_0=\frac{\pi^2}{3}(\frac{k_B}{e})^2=2.44\times 10^{-8} ~\Omega$WK$^{-2}$ at exact zero $T$, because two pondering functions summarize at the same Fermi energy, $\mu$. At finite $T$, for example, if $\tilde{\sigma}_{zx}(\xi)$ is a constant or a linear function of $\xi$, $L^A_{zx}$ also remains $L_0$.

Here, we summarize general rules about the anomalous WF law. At zero temperature, $L^A_{zx}=L_0$ is valid, as long as the total Berry curvature is smooth to energy, which is commonly true. At finite $T$, $L^A_{zx}$ is close to $L_0$ if $\tilde{\sigma}_{zx}(\xi)$ has an approximately antisymmetric profile with respect to $\mu$ in the energy window of $\sim 5 k_B T$, because two pondering factors contribute the same summation of the Berry curvature in this case. Otherwise, $L^A_{ij}$ can vary strongly from $L_0$. Such a violation of the anomalous WF law  is caused by the distribution of the Berry curvature, instead of the extrinsic scattering.

The total Berry curvature $\tilde{\sigma}_{zx}(\xi)$ is determined by the intrinsic band structure of a give material.
It is not surprising that $L^A_{zx}$ depends sensitively on the position of $\mu$  because $\kappa_{zx}^A(\mu)$ and $\sigma_{zx}^A(\mu)$ do. In experiment, Mn$_3$Ge(Sn) is usually off-stoichiometric in the form of Mn$_{3+x}$Ge(Sn)$_{1-x}$ \cite{Li2017,Nakatsuji2015,Nayak2016,Kiyohara2016}, where there are usually excess of Mn but deficiency of Ge(Sn). Such an off-stoichiometry can shift the chemical potential up in energy compared to the charge neutral point.
%In addition, such a disorder also smooths the Berry curvature dependence on energy, which tends to recover $L_0$.

By shifting $\mu$ slightly above, we can successfully reproduce the general trend of $T$-dependence of $L^A_{zx}$ for both Ge and Sn compounds in Fig.~\ref{calculation-Lxz}. For Mn$_3$Ge at $\mu=180$ meV above the charge neutral point, $L^A_{zx}$ drops quickly from 100 $\sim$ 250 K and goes up again after 250 K, qualitatively consistent with the experiment. The first drop is induced by dips of $\sigma^A_{zx}$ at $\sim$0.1 and $\sim$0.3 eV, because the $\kappa^A_{zx}(T)$ sums smaller Berry curvature than $\sigma^A_{zx}(T)$ does in this case (see Eqs.\ref{simga}-\ref{kappa}). We note that the dip at $\sim$0.1 eV is induced by an anti-crossing gap while the dip at $\sim$0.3 eV is caused by a Weyl point. (See the supplementary Fig.~\ref{fig:dip-band})
The following up-turn is related to the increase of $\sigma^A_{zx}$ after these dips. We also note that the theoretical violation is smaller than the experimental one. This may point to a significant role played by electronic correlations, which is neglected in the density-functional theory employed here. For Mn$_3$Sn at $\mu=140$ meV, however, $L^A_{zx}$ remains close to $L_0$ in a large temperature range.

Different behaviors between two materials originate in their different Berry curvature $\tilde{\sigma}_{zx}(\xi)$, as shown in Fig.~\ref{calculation-Lxz}(a) and (c). For Mn$_3$Sn, $\tilde{\sigma}_{zx}(\xi)$ is approximately anti-symmetric with respect to $\mu=140$ meV in an energy window of nearly 200 meV, inducing $L^A_{zx} \approx L_0$ up to 400 K and even above. However, $\mu$ may vary from such an ideal anti-symmetric point in different samples. This explains the observed deviation of of $L^A_{zx}$ in a different Mn$_3$Sn sample, as shown in Fig. ~\ref{S9}. To demonstrate the sensitive role of the chemical potential, we show $L^A_{zx}-T$ curves for different $\mu$ in Fig.~\ref{calculation-Lxz}.
Because it depends on the competition of the Berry curvature [$\tilde{\sigma}_{zx}(\xi)$] profile and the temperature ($5 k_B T$), $L^A_{zx} \approx L_0$ at low $T$ for different $\mu$.
% Furthermore, our choice of the chemical potential to simulate the experimental $L_{zx}$ is also supported by an independent measurement on the anomalous Nernst coefficient, $\alpha^A_{zx}$.
% At low temperature, the Mott relation holds,
% \begin{equation}
% \frac{\alpha_{zx}^A}{T} |_{T\rightarrow 0} = -\frac{\pi^2k_B^2}{3|e|} \frac{d\sigma^A_{zx}}{d\mu},
% \label{theory-Mottrelation}
% \end{equation}
% At the chosen $\mu$ (180 meV for Mn$_3$Ge and 140 meV for Mn$_3$Sn), $\frac{d\sigma^A_{zx}}{d\mu}$ is negative and positive for Ge and Sn compounds, respectively, according to Fig.~\ref{calculation-Lxz}.
% \textcolor{red}{
% Therefore, $\frac{\alpha_{zx}^A}{T}$ presents positive and negative signs for Mn$_3$Ge and Mn$_3$Sn, respectively, which is well consistent with our experiment and Ref.~\onlinecite{Ikhlas2017}. [please confirm this point]}
In addition, there are interesting topological features near the chose chemical potential (see Fig.~\ref{calculation-Lxz}e-f). Along the $\Gamma-M$ line in the Brillouin zone, there are two Weyl points induced by the crossing between the lowest and second lowest conduction bands. These Weyl points were not revealed in previous study which usually focuses on the crossing between the lowest conduction band and the highest valence band. For Mn$_3$Sn at $\mu=140$ meV, a Weyl point exists and contributes large Berry curvature to the anomalous Hall conductivity. For Mn$_3$Ge at $\mu=180$ meV, however, the Weyl cone is strongly tilted so that negative and positive Berry curvatures nearly cancel each other near the Weyl point (see more information in the supplementary Fig.~\ref{fig:band}).

% The Berry curvature usually changes dramatically at the band edge. Therefore, the total Berry curvature may vary quickly near a gap or a Weyl point. Besides the anti-crossing gap illustrated in Fig.~\ref{fig:band}, Weyl points in Mn$_3$Ge are closer to the charge neutral point than those in Mn$_3$Sn (see Ref.\cite{Yang2017}). Therefore, from the charge neutral point to around 150 meV above (approximately $5 k_B T$ of the room temperature), $\tilde{\sigma}_{zx}$ is smoother to energy in Mn$_3$Sn than that in Mn$_3$Ge, as shown in Fig.~\ref{calculation-Lxz}. \color[rgb]{1,0,0}This is the general reason for the violation or validity of the anomalous transverse WF law  at finite temperature. Accordingly, changing the Mn content may lead to an evolution of the finite temperature magnitude of the Lorenz ratio as suggested by a comparison of various data sets in Mn$_3$Sn (See the supplement\cite{supplement}).  We also note that the theoretical violation is smaller than the experimental one. This may point to a significant role played by electronic correlations, which is neglected in the non-interacting theory employed here.
% \color{black}

\section*{Concluding remarks}
To summarize, we measured  counterparts of the anomalous Hall effect associated with the flow of entropy and found that the Wiedemann-Franz law linking the magnitude of the thermal and electrical Hall effects is valid at zero-temperature, but a finite deviation emerges above 100 K. This deviation is not caused by inelastic scattering, but arises because of a mismatch in thermal and electrical summations of Berry curvature over the Fermi surface. The Bridgman relation, which links anomalous Nernst and Ettingshausen coefficients is satisfied over the whole temperature range. Finally, we observed that the room-temperature $\alpha^A_{zx}/\sigma^A_{zx}$ ratio is close to $k_B/e$.\\

\textbf{METHODS}\\
\textbf{Sample preparation and transport measurement:} Single crystals of Mn$_{3}$Ge were grown from polycrystalline samples, using Bridgman-Stockbarger technique. The raw materials, Mn (99.99\% purity) and Ge (99.999\% purity), were weighed and mixed in an Argon glove box with a molar ratio of 3.3:1, loaded in an alumina crucible then sealed in a vacuum quartz ampule. The mixture was heated up to 1050 $^{o}$C, remained  for 2 hours to ensure homogeneity of melt, then was cooled slowly down to 800 $^{o}$C to obtain polycrystalline samples. The polycrystalline Mn$_{3}$Ge were ground, loaded in an alumina crucible and sealed into another vacuum quartz ampule. The growth temperature was controlled at 980 $^{o}$C and 800 $^{o}$C for high-temperature and low-temperature end, respectively. Finally, to obtain high temperature hexagonal phase, the quartz ampule was quenched with water. The single crystals were cut by a wire saw into typical dimensions of $0.3\times1.5\times2$ mm$^3$ for transport measurements. The stoichiometry  was found to be Mn$_{3.08}$Ge$_{0.92}$ (Mn:Ge = 3.32-3.35:1) using energy dispersive X-ray spectroscopy (EDX), This is close to the ratio of the raw materials and comparable to previous reports\cite{Kiyohara2016}.

Longitudinal and Hall resistivity were measured by the standard four-probe method using a current source (Keithley 6221) with a DC-Nanovoltmeter (Keithley 2182A) in a commercial measurement system (PPMS, Quantum Design). The thermal conductivity and thermal Hall effect were performed using a heater and two pairs of thermocouples in the PPMS in a high-vacuum environment\cite{Li2017}. For temperatures below 4.2 K, the measurements were performed in a dilution refrigerator inserted in a 14 T superconducting magnet using  one-heater-three-thermometers set-up, allowing to measure longitudinal and transverse transport coefficients with the same contacts.

\textbf{Theoretical calculations:} The band structure was calculated with the density-functional theory in the framework of the generalized-gradient approximation~\cite{Perdew1996}. The Bloch wave functions were projected to atomic-orbital-like Wannier functions~\cite{Mostofi2008}. Based on the Wannier-projected tight-binding Hamiltonian,  we calculated the Berry curvature and the anomalous Hall conductivity in the clean limit. More details can be found in Ref. ~\cite{Zhang2017}. As shown in Fig.~\ref{calculation-Lxz}(a) and (c), the $\tilde{\sigma}_{zx}(\xi)$ data are analytically fitted so that integrals in Eqs.~\ref{simga} and \ref{kappa} can be evaluated accurately with dense energy steps.

% Besides the spin structure shown in Fig. 6a, we also investigated different spin structures by rotating all spins in the \textit{xy} plane. Different spin structures support the same conclusion on the Berry curvature hot spots and anti-crossing gaps  in both compounds. We note that the AHE vanishes without including SOC even if the noncolinear AFM appears, because the spin-rotational symmetry appears in this case.

\textbf{Data availability:} The data that support the findings of this study are available  upon reasonable request.

\textbf{Acknowledgements} We thank Alaska Subedi for numerous discussions. This work was supported by the National Science Foundation of China (Grant No. 51861135104 and No. 11574097) and The National Key Research and Development Program of China(Grant No.2016YFA0401704). Z. Z. was supported by the 1000 Youth Talents Plan. K. B. was supported by China High-end foreign expert program and Fonds-ESPCI-Paris and by the KITP through the National Science Foundation Grant No. NSF PHY17-48958.  B.Y. acknowledges the financial support by the Willner Family Leadership Institute for the Weizmann Institute of Science,
the Benoziyo Endowment Fund for the Advancement of Science, Ruth and Herman Albert Scholars Program for New Scientists, and the European Research Council (ERC) under the European Union¡¯s Horizon 2020 research and innovation programme (grant agreement No. 815869). B.Y. thanks the fruitful discussion with Yang Zhang at the Max Planck Institute in Dresden.

%\noindent
%* \verb|binghai.yan@weizmann.ac.il|\\
%* \verb|zengwei.zhu@hust.edu.cn|\\
%* \verb|kamran.behnia@espci.fr|\\

\clearpage
% Add 'S' to the numbering inside the supplement
\renewcommand{\thesection}{S\arabic{section}}
\renewcommand{\thetable}{S\arabic{table}}
\renewcommand{\thefigure}{S\arabic{figure}}
\renewcommand{\theequation}{S\arabic{equation}}

\setcounter{section}{0}
\setcounter{figure}{0}
\setcounter{table}{0}
\setcounter{equation}{0}
{\large\bf Supplemental Material for ``Finite-temperature violation of the anomalous transverse Wiedemann-Franz law in absence of inelastic scattering''}

\setcounter{figure}{0}
%\section{Samples and stoichiometry}

\begin{figure}[hbp]
\includegraphics[width=8cm]{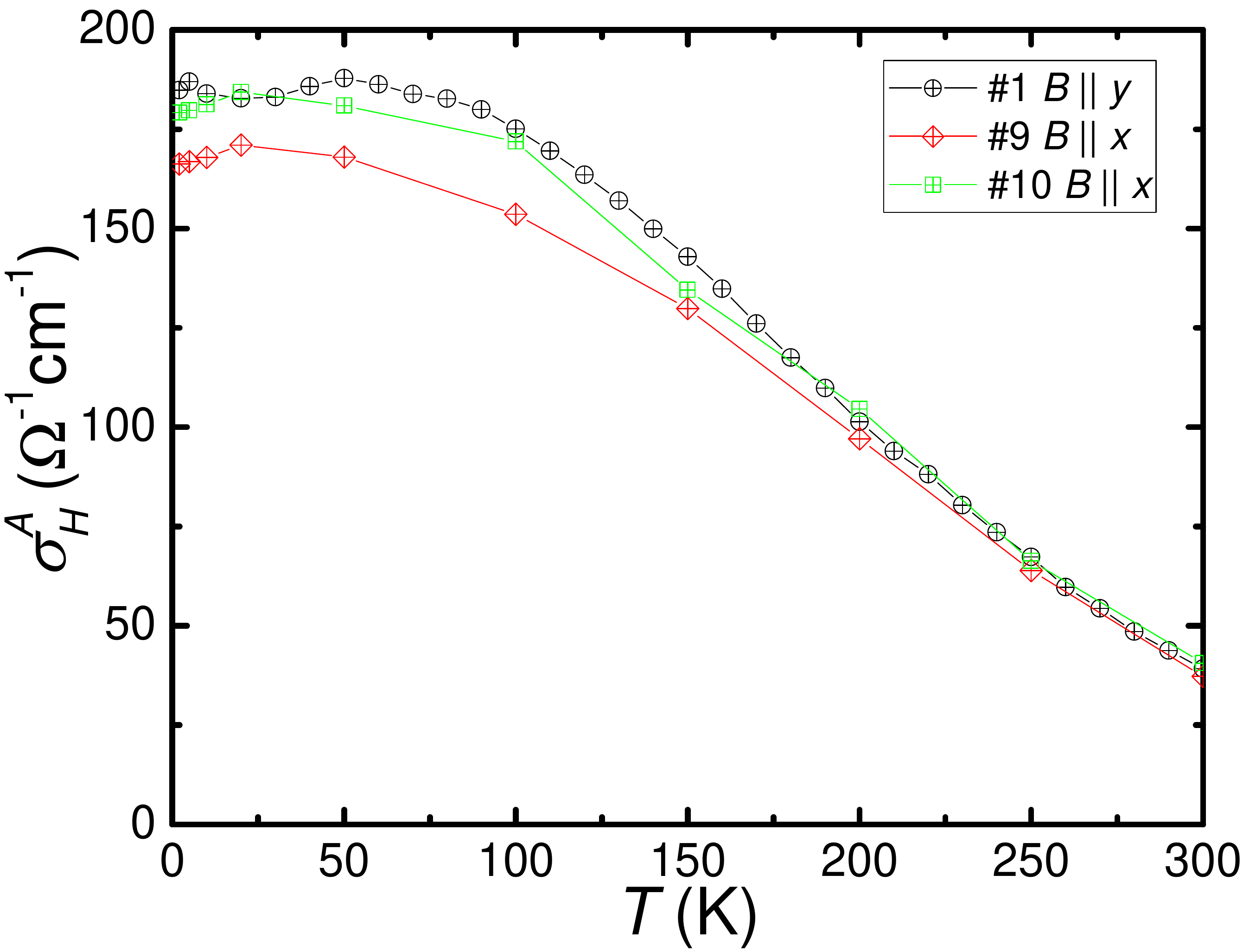}
\caption{Temperature dependence of anomalous Hall conductivity in three samples from the same batch in two configurations,  $B \parallel y$ and $B \parallel x$.}\label{S7}
\end{figure}
\begin{figure}
\includegraphics[width=8cm]{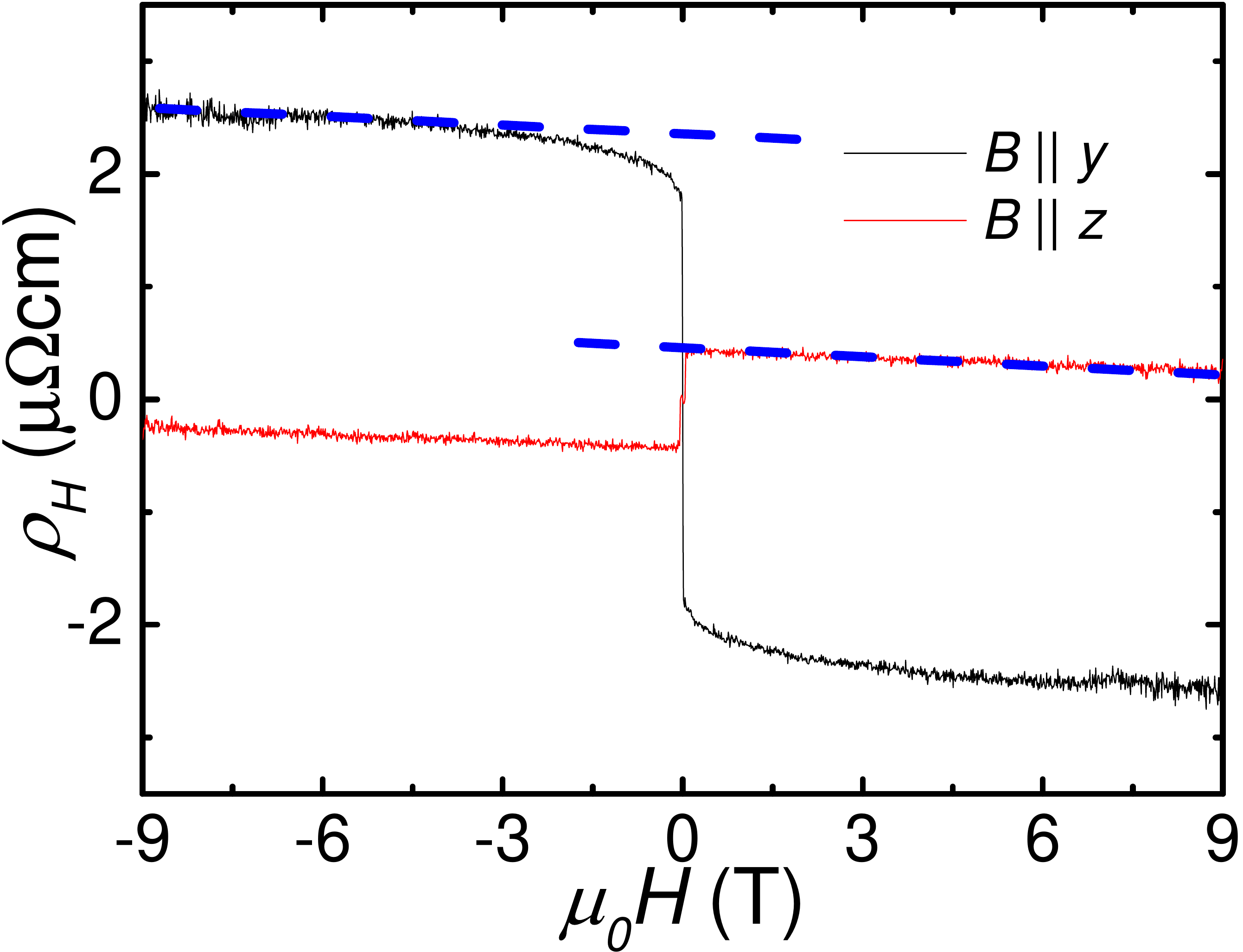}
\caption{Hall resistivity $\rho_H$ vs. $B$ for $B \parallel y$ (black line) and $B \parallel z$ (red line) at 2 K. The blue lines are linear fits to quantify the ordinary Hall resistivity.}\label{S8}
\end{figure}
\section{Magnitude of the anomalous Hall conductivity}
The zero-temperature anomalous Hall conductivity found here (190 $\Omega^{-1}$cm$^{-1}$) is to be compared with previous reports of 150-450 $\Omega^{-1}$cm$^{-1}$ in Ref.~\cite{Nayak2016_S} and 320-380 $\Omega^{-1}$cm$^{-1}$ in Ref.~\cite{Kiyohara2016_S}. The difference in stoichiometry between the present sample (Mn$_{3.1}$Ge$_{0.9}$) and the one studied in Ref.~\cite{Kiyohara2016_S} (Mn$_{3.05}$Ge$_{0.95}$) combined to the fact that the magnitude of the anomalous Hall resistivity depends on Mn content\cite{Kiyohara2016_S}, provide an explanation for this apparent discrepancy. The temperature dependent anomalous Hall conductivity in different samples cut from the same batch are similar (see Fig. \ref{S7}).
\section{Carrier density}
The field dependence of Hall resistivity along two different directions, yield the same value for $\partial \rho_H / \partial B$ at  high magnetic field  at 2 K (see Fig.\ref{S8}). The slope was found to be $R_H= \partial \rho_H/ \partial B = -0.02~\mu\Omega\,$cm/T. Using a single-band model ($n=\frac{1}{{\rm e}R_H}$), we found a carrier density of  3.1$\times$10$^{22}$ cm$^{-3}$, which is comparable  to Ref. \cite{Nayak2016_S}. A small anomalous Hall resistivity observed previously\cite{Nayak2016_S} in the $B \parallel z$ configuration is also observed in our sample.

\begin{figure}
\includegraphics[width=8cm]{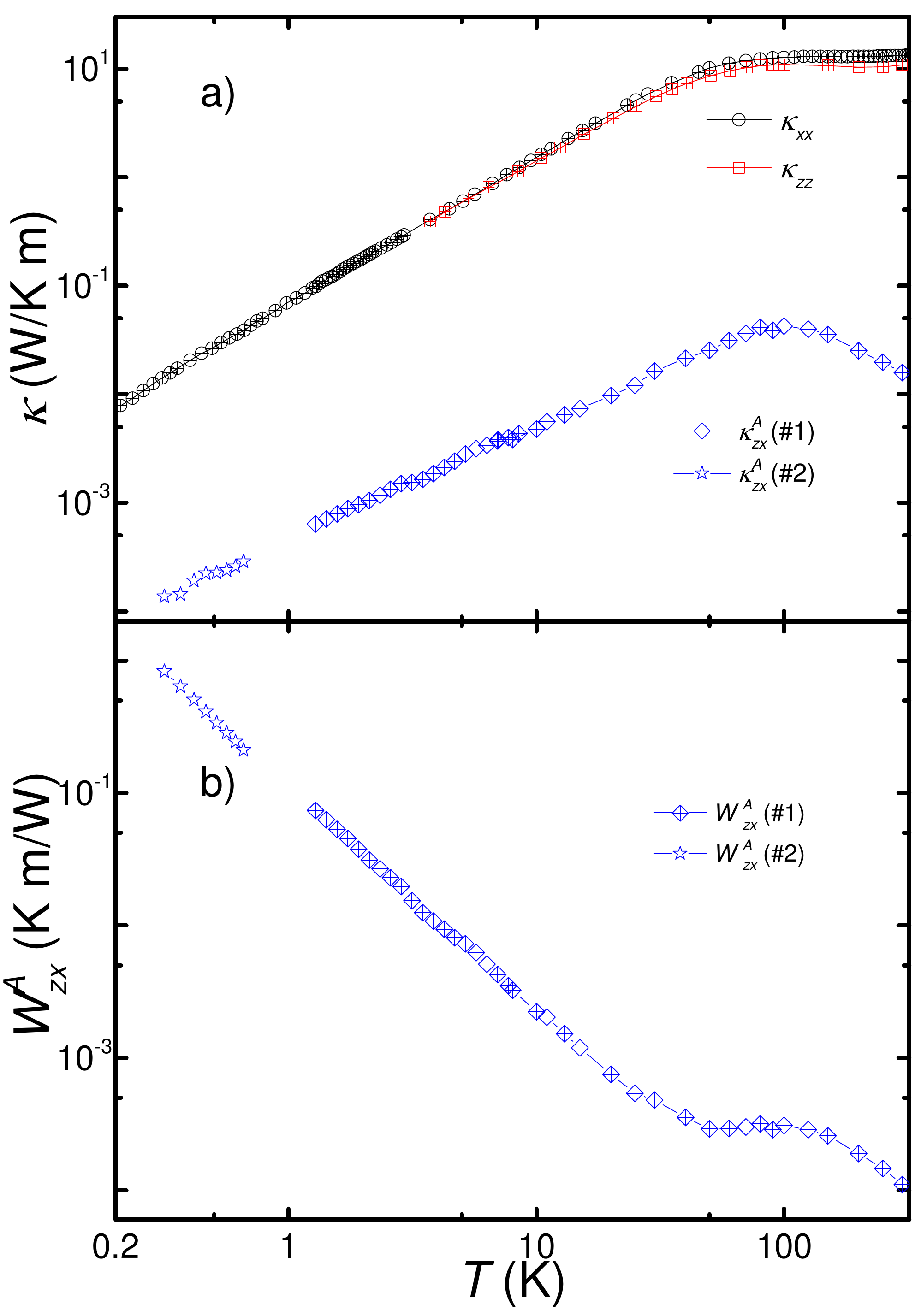}
\caption{ a) Temperature dependent longitudinal thermal conductivity $\kappa_{xx}$ and $\kappa_{zz}$ and anomalous transverse thermal conductivity $\kappa^A_{zx}$, the diamond and star denote the data measured in \#1 and \#2, respectively. b) The anomalous transverse thermal resistance $W^A_{zx}$ with the function of temperature.}\label{S1}
\end{figure}

\section{Longitudinal thermal transport}
\begin{figure}
\includegraphics[width=8cm]{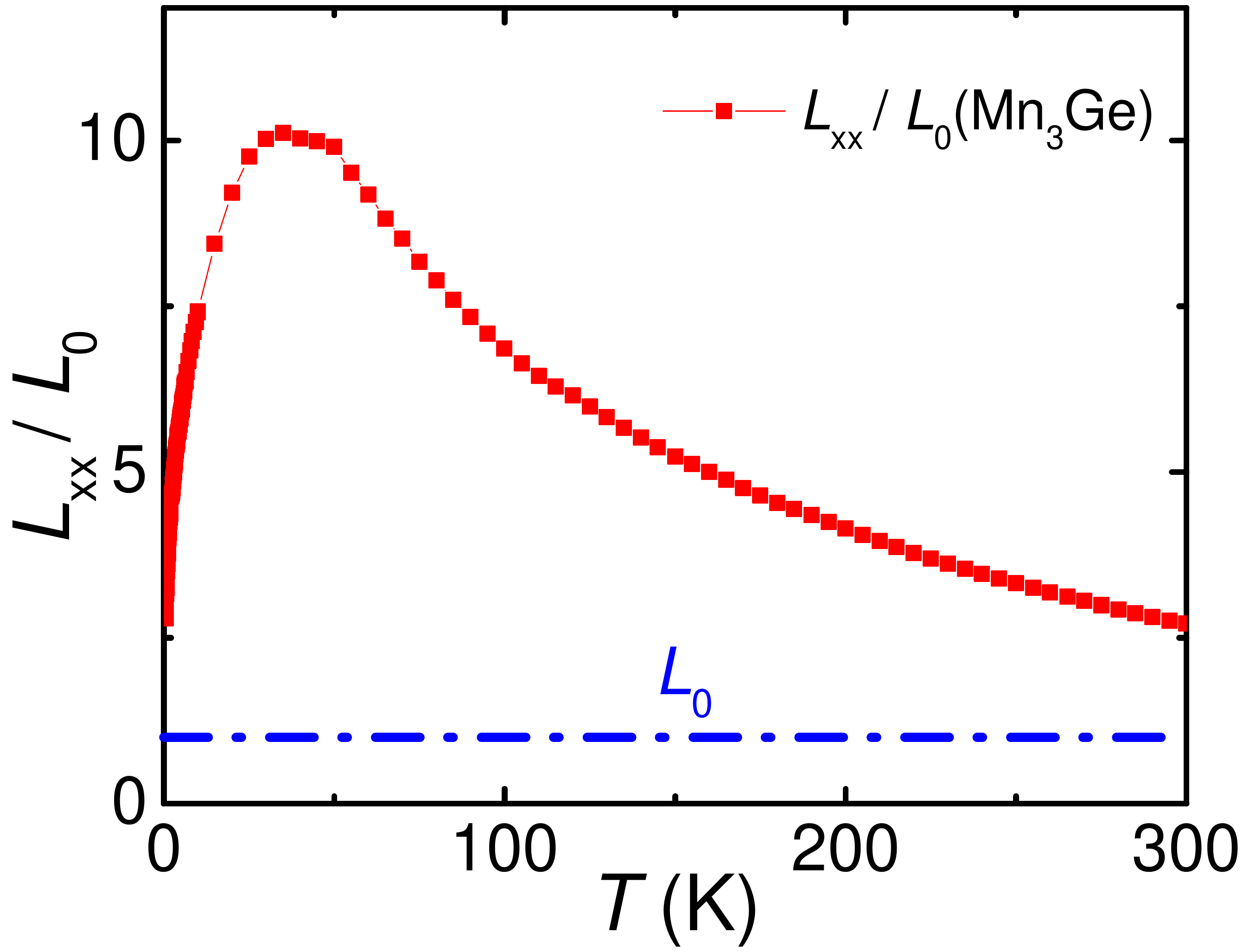}
\caption{The Lorenz ratio for longitudinal thermal and electrical transport in Mn$_3$Ge. }\label{L_xx}
\end{figure}

Fig. \ref{S1} shows the temperature dependence of thermal conductivity along $x$ axis, $\kappa_{xx}$, and along $z$ axis, $\kappa_{zz}$ as well as the anomalous transverse thermal conductivity, $\kappa^A_{zx}$. The latter was extracted using the anomalous thermal resistivity, $W^A_{zx}$ and the longitudinal conductivities. Fig.\ref{L_xx} shows Lorenz ratio $L_{xx}$ for longitudinal transport in Mn$_3$Ge. At finite temperature, $L_{xx}/L_{0}$ exceeds unity indicating a large contribution from phonons. As expected, the ratio tends to unity in the zero temperature limit.

\section{The Kelvin Relation}
\begin{figure}[b]
\includegraphics[width=7.5cm]{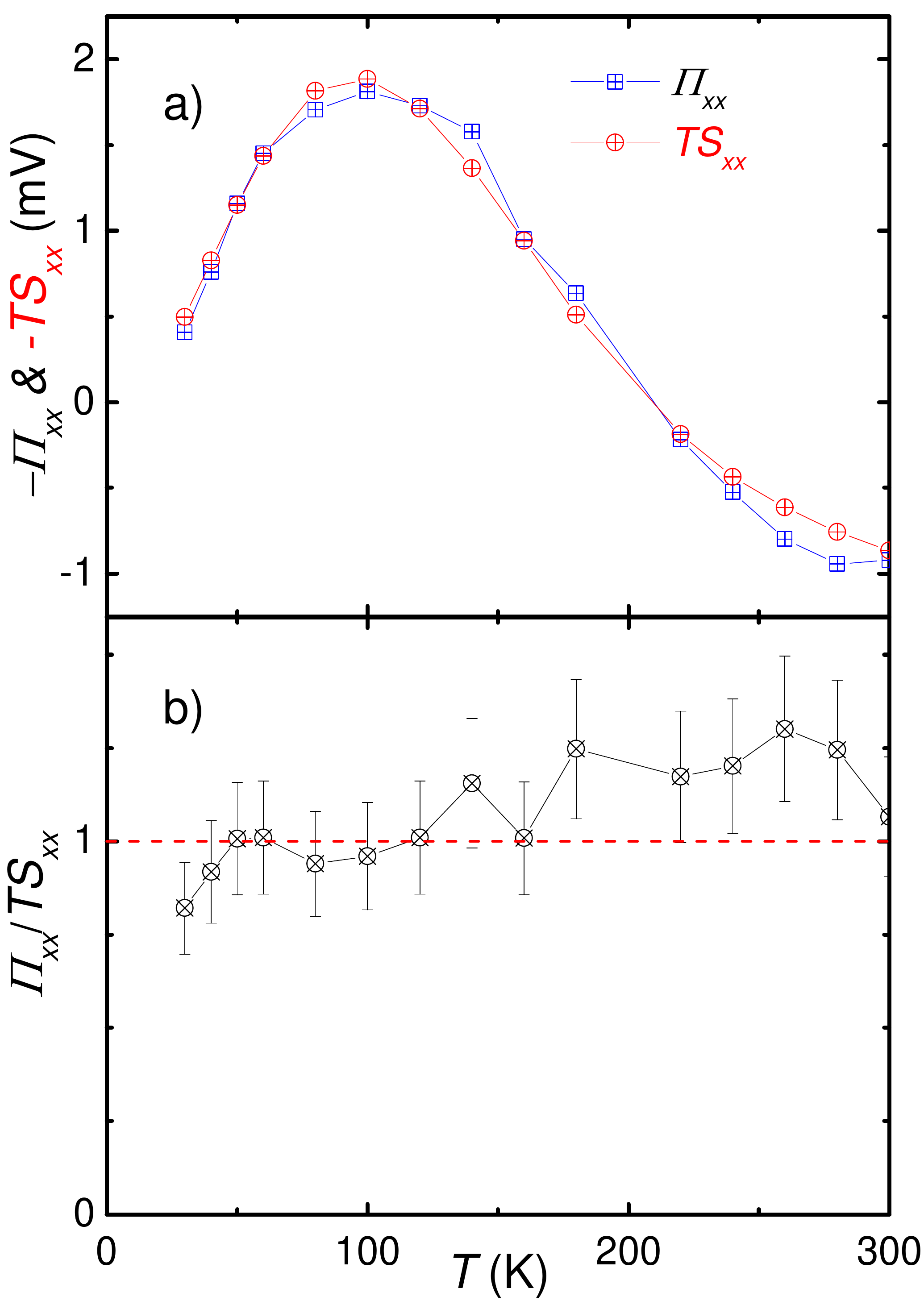}
\caption{\textbf{Kelvin relation:} a) The comparison of Peltier coefficient, $\Pi_{xx}$ and Seebeck, $S_{xx}$ times temperature, $T$ at different temperatures. b) The ratio of $\Pi_{xx}/TS_{xx}$ as a function of temperature.}\label{S4}
\end{figure}

The  Peltier effect is the heat current produced by a charge current. The Kelvin relation relates the Peltier and Seebeck coefficients: $\Pi_{xx}= T S_{xx}$. Fig. \ref{S4}b shows  $\Pi_{xx}$ and $T S_{xx}$ in Mn$_{3}$Ge. The ratio of $\Pi_{xx} / T S_{xx}$ is close to 1 in the whole temperature range. The  Peltier coefficient $\Pi_{xx}$ and the Ettingshausen coefficient $\epsilon_{zx}$ were obtained by symmetrizing the signals obtained by injecting positive and negative currents through the sample in order to eliminate the Joule heating effect. Comparing to Ettingshausen effect, we can measure the heat absorb/release from the sample. We measured the temperature gradient along the sample to get the heat combining with the thermal conductivity\cite{Garrido_S}.

\section{Thermal evolution of anomalous transverse response}
\begin{figure}
\includegraphics[width=7.5cm]{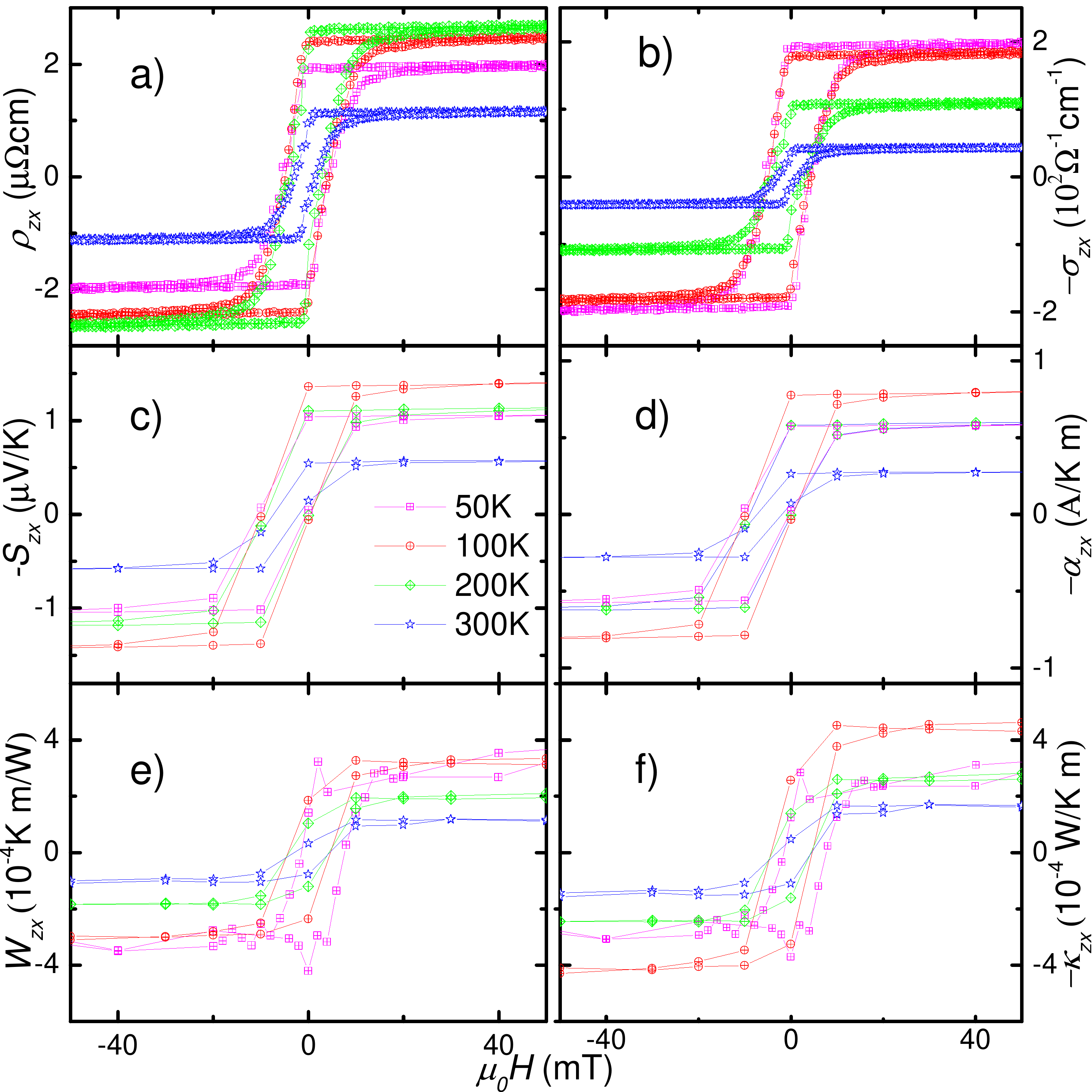}
\caption{  Field dependence of transverse transport coefficients at different temperatures.}\label{S5}
\end{figure}

Fig. \ref{S5} displays the field dependence of the three transverse coefficients at different temperatures. The temperature dependence of the anomalous transverse transport coefficients presented in main text were deduced from these curves.

\section{Specific heat C$_P$ and magnetic susceptibility $\chi$}

\begin{figure}[b]
\includegraphics[width=6cm]{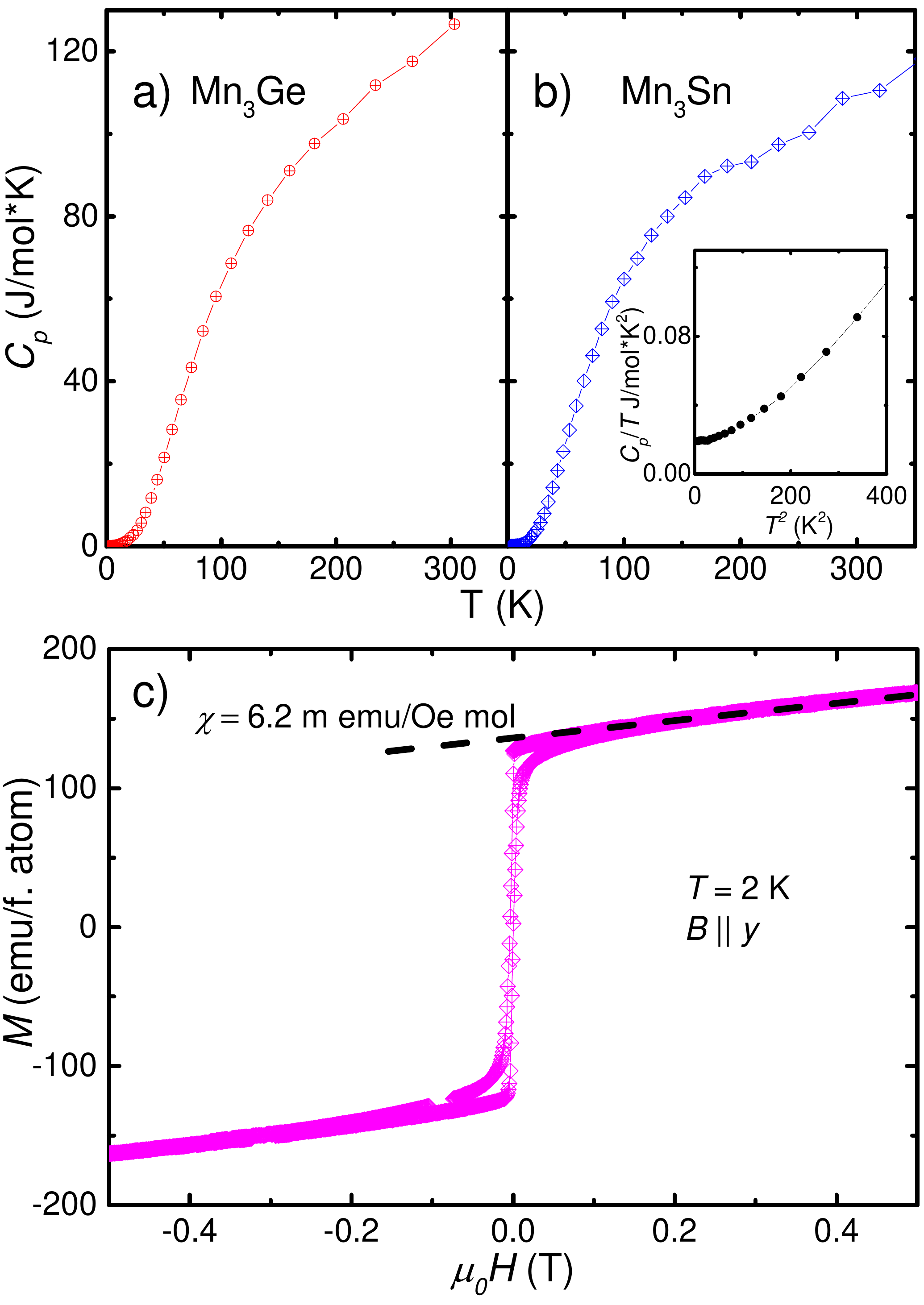}
\caption{Temperature dependence of specific heat of Mn$_3$Ge, a) and Mn$_3$Sn ,b). Inset show the $C/T$ $vs.$ $T^2$ of Mn$_3$Sn . c) Magnetization curve of Mn$_3$Ge measured at 2 K, the susceptibility $\chi$ = 6.2 $\times$ 10$^{-3}$ emu.Oe$^{-1}$mol$^{-1}$ is extracted from the high-field slope (black dashed line).}\label{S3}
\end{figure}

\begin{figure}[t]
\includegraphics[width=8cm]{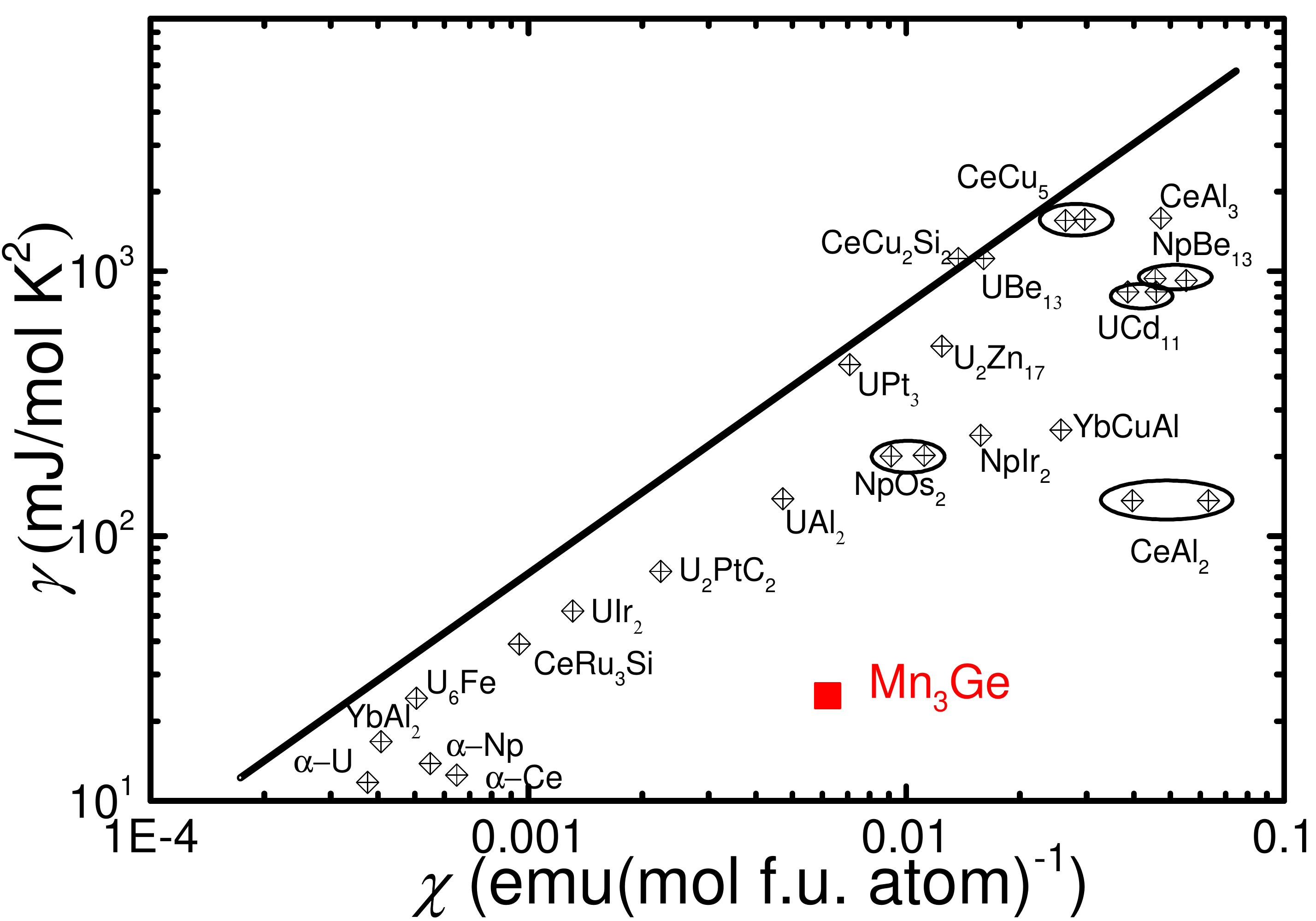}
\caption{ The electronic specific heat vs. magnetic susceptibility $\chi$ ; the red rectangle refers to Mn$_3$Ge, the others are from Ref. \cite{Coleman2007_S}, the solid line represent $\frac{\pi^2\kappa^2_B}{3\mu^2_B}\chi$. The large magnetic susceptibility of Mn$_3$Ge is not caused by correlated electrons but by ordered spins.}\label{S6}
\end{figure}

\begin{figure}[t]
\includegraphics[width=7.5cm]{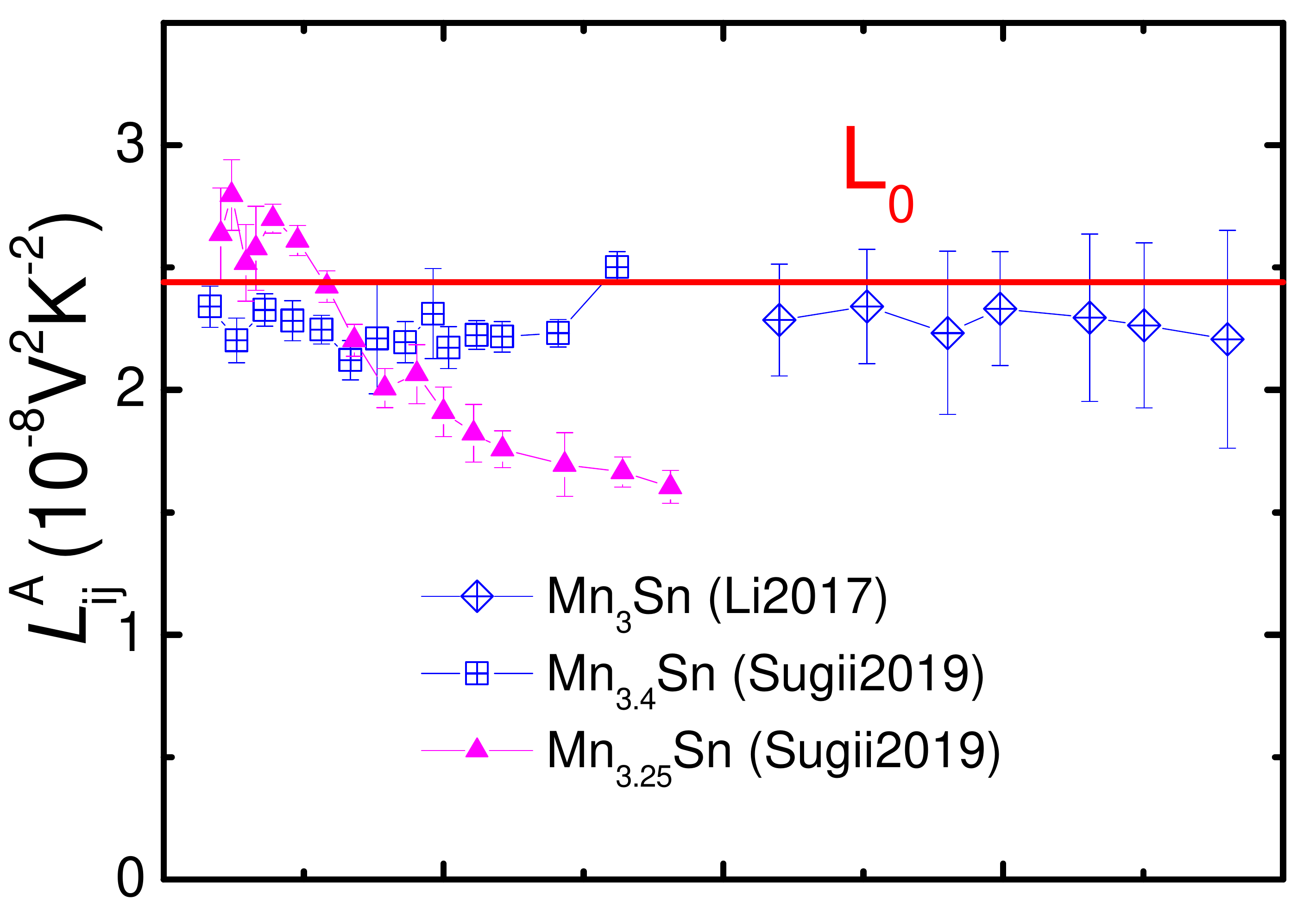}
\caption{ The Lorentz ratio in Mn$_3$Sn with different content of Mn.(digitized from\cite{Sugii_S,Li2017_S})}\label{S9}
\end{figure}

The Fig. \ref{S3} a) and b) show the specific heat  data for Mn$_3$Ge and Mn$_3$Sn. As seen in the inset,  the electronic contribution in Mn$_3$Sn is 19.1 mJ mol$^{-1}$K$^{-2}$.  From the low-temperature magnetization curve (Fig. \ref{S3}c),  the magnetic susceptibility  $\chi$ is deduced to be  6.2 $\times$ 10$^{-3}$ emu Oe$^{-1}$mol$^{-1}$. As seen in Fig. \ref{S6}, the magnetic susceptibility is much larger than what is expected for the Pauli susceptibility of mobile electrons according to the  Wilson ratio. This implies that most of the susceptibility is due to Mn spins.

\section{The dependence of Lorenz ratio in $\rm{Mn}_3\rm{Sn}$ on stoichiometry}

The focus of this paper is the validity and/or the violation of the anomalous transverse WF law in Mn$_3$Ge. Nevertheless, let us notice that in the case of   Mn$_3$Sn, the published data suggests that the the precise chemical composition of the compound can effect validity or verification of the WF law with little effect on the temperature dependence of resistivity. The precise location of the chemical potential depends on the the Mn. A violation of the WF law in  Mn$_3$Sn is conceivable if the chemical potential shifts away from the $\mu=140$ meV  (See Fig.\ref{calculation-Lxz}).

%\clearpage

\section{The band structure and Berry curvature}
\begin{figure*} [b]
\includegraphics[width=14cm]{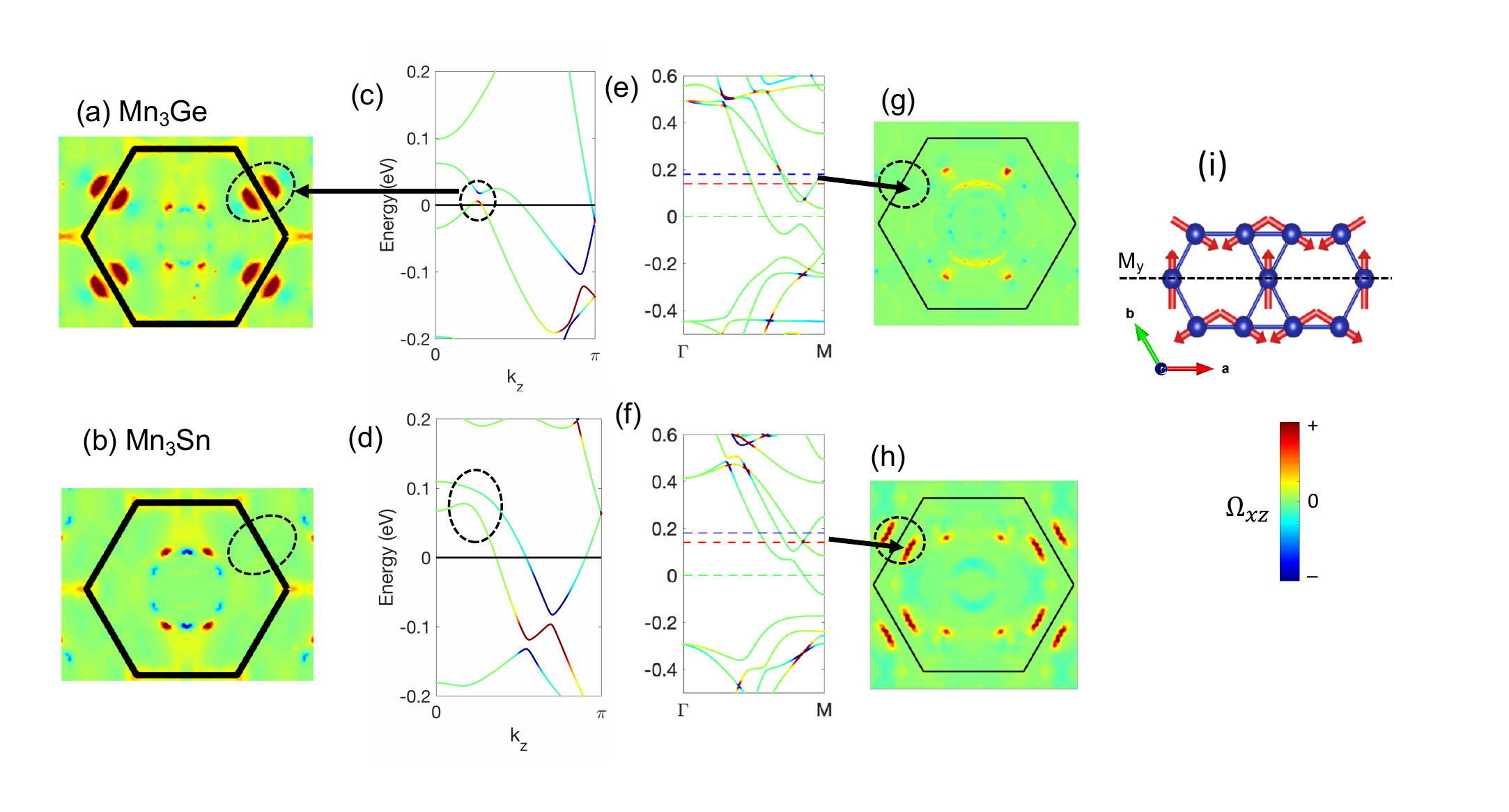}
\caption{\textbf{ Contrasting Mn$_3$Ge and Mn$_3$Sn in the band structure and Berry curvature.} The upper and lower panels are for Mn$_3$Ge and Mn$_3$Sn, respectively.
(a) and (b) The Berry curvature $\Omega_{xz}$ distribution in the $k_x k_y$ plane integrated along $k_z$ for Mn$_3$Ge and Mn$_3$Sn, respectively. The color bar of $\Omega_{xz}$ is in arbitrary unit. Corresponding noncollinear AFM spin structure is shown in (i). There is a $M_y$ mirror symmetry to constrain that only $\Omega_{xz}$ is even to the mirror plane, i.e. only the anomalous Hall conductivity $\sigma_{xz}$ is nonzero. The dashed circle in (a) indicates two hot spots near the $M$-point in the Brillouin zone (the black hexagon). (c)-(d) The band structure along $k_z$ through the center of a hot spot for Mn$_3$Ge and Mn$_3$Sn, respectively. The Fermi energy is set to zero. The color code indicates the Berry curvature distribution in the band structure. At the charge neutral point, the Berry curvature is dominantly contributed by an anti-crossing gap in along the $k_z$ axis for Mn$_3$Ge, while such a gap is missing for Mn$_3$Sn.
(e)-(f) The band structure along $\Gamma-M$. When it lies at 180 meV, the chemical potential crosses a type-II Weyl point for Mn$_3$Ge. Because of the strong tilting of the Weyl cone, its net contribution to the total Berry curvature (sum over all occupied bands) is small. In contrast, he chemical potential of 140 meV crosses a weakly-tilted Weyl point for Mn$_3$Sn, inducing large Berry curvature.
\label{fig:band}}
\end{figure*}

\begin{figure*}[b]
\centering
\includegraphics[width=13cm]{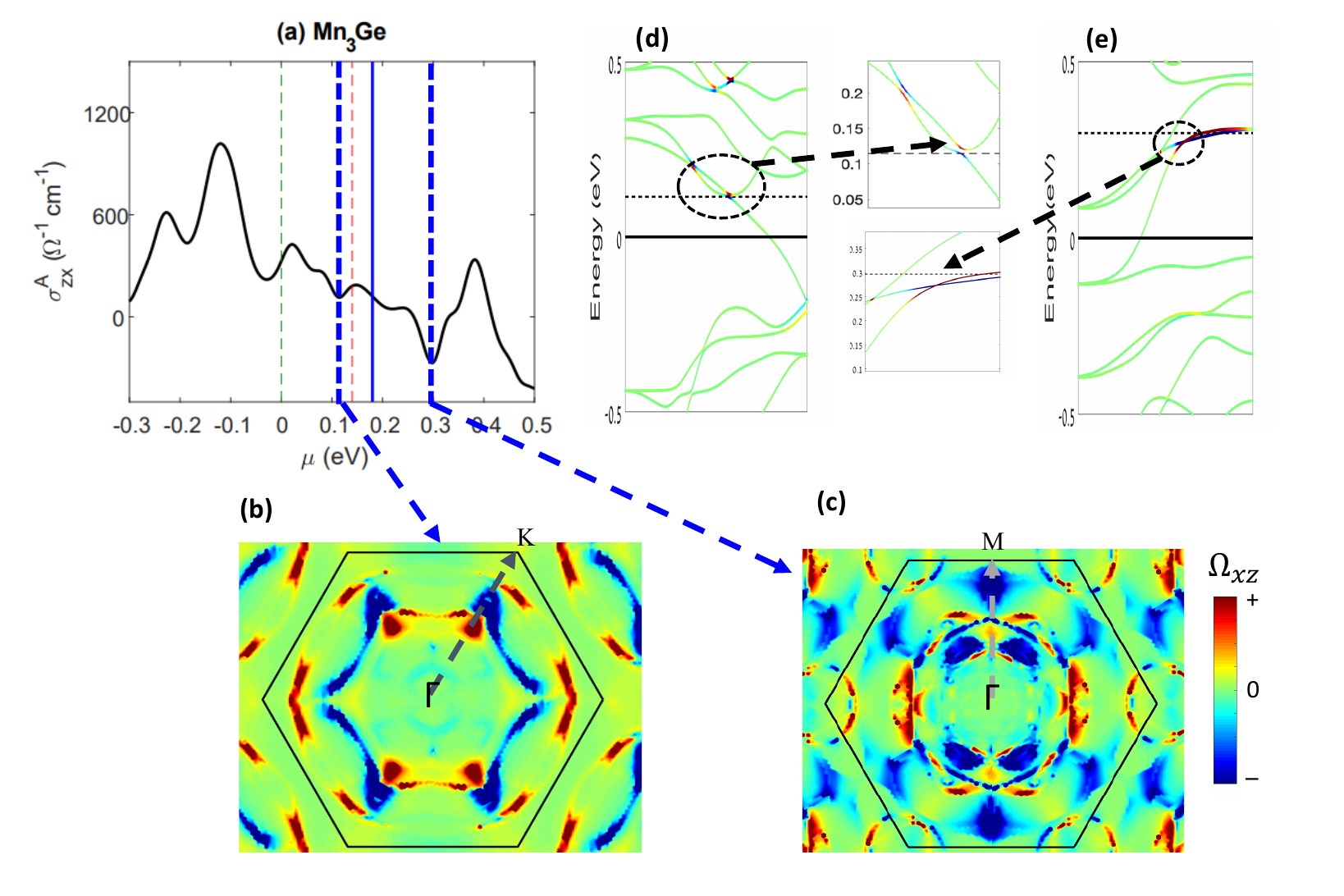}
\caption{\textbf{ Band structure and Berry curvature of Mn$_3$Ge.}
(a) The anomalous Hall conductivity. We put Fig.~\ref{calculation-Lxz}a here for the completeness. (b) The anomalous Hall conductivity distribution corresponding to the dip at $\mu=0.115$ eV.  (d) Along the $\Gamma-K$ direction but with $k_z=0.14$ $(2\pi /c)$ ($c$ is the lattice parameter along the $z$ direction), the negative Berry curvature is contributed by an anti-crossing gap.  (c) The anomalous Hall conductivity distribution corresponding to the dip at $\mu=0.297$ eV. (e) Along the $\Gamma-M$ direction but with $k_z=0.3867$ $(2\pi /c)$ ($c$ is the lattice parameter along the $z$ direction), the negative Berry curvature is contributed by a type-II Weyl point.
\label{fig:dip-band}}
\end{figure*}

\end{document}